\documentclass[sigconf]{acmart}

\usepackage{booktabs}
\usepackage{algorithm}
\usepackage{algpseudocode}
\usepackage{bbold}
\usepackage{enumitem}
\usepackage{subfigure}
\usepackage{colortbl}
\usepackage{balance}

\newcommand{\todo}[1]{\textcolor{red}{#1}}

\newcommand{\sz}[1]{\textcolor{blue}{#1}}

\newtheoremstyle{mydef}
{2ex}
{2ex}
{\itshape}
{}
{\scshape}
{: }
{0.5em}
{}
\theoremstyle{mydef}

\copyrightyear{2018}
\acmYear{2018} 
\setcopyright{iw3c2w3}
\acmConference[WWW 2018]{The 2018 Web Conference}{April 23--27, 2018}{Lyon, France}
\acmBooktitle{WWW 2018: The 2018 Web Conference, April 23--27, 2018, Lyon, France}
\acmPrice{}
\acmDOI{10.1145/3178876.3186067}
\acmISBN{978-1-4503-5639-8/18/04} 

\begin{document}

\title{Ad Hoc Table Retrieval using Semantic Similarity}

\author{Shuo Zhang}
\affiliation{%
  \institution{University of Stavanger}
}
\email{shuo.zhang@uis.no}

\author{Krisztian Balog}
\affiliation{%
  \institution{University of Stavanger}
}
\email{krisztian.balog@uis.no}

\begin{abstract}
We introduce and address the problem of ad hoc table retrieval: answering a keyword query with a ranked list of tables.  This task is not only interesting on its own account, but is also being used as a core component in many other table-based information access scenarios, such as table completion or table mining.  The main novel contribution of this work is a method for performing semantic matching between queries and tables.  Specifically, we (i) represent queries and tables in multiple semantic spaces (both discrete sparse and continuous dense vector representations) and (ii) introduce various similarity measures for matching those semantic representations.  We consider all possible combinations of semantic representations and similarity measures and use these as features in a supervised learning model.  Using a purpose-built test collection based on Wikipedia tables, we demonstrate significant and substantial improvements over a state-of-the-art baseline.
\end{abstract}



\maketitle

\section{Introduction}

Tables are a powerful, versatile, and easy-to-use tool for organizing and working with data.  Because of this, a massive number of tables can be found ``out there,'' on the Web or in Wikipedia, representing a vast and rich source of structured information.  Recently, a growing body of work has begun to tap into utilizing the knowledge contained in tables.  A wide and diverse range of tasks have been undertaken, including but not limited to 
(i) searching for tables (in response to a keyword query~\citep{Cafarella:2008:WEP,Cafarella:2009:DIR,Venetis:2011:RST,Pimplikar:2012:ATQ,Balakrishnan:2015:AWP,Nguyen:2015:RSS} or a seed table~\citep{DasSarma:2012:FRT}), 
(ii) extracting knowledge from tables (such as RDF triples~\citep{Munoz:2014:ULD}), and
(iii) augmenting tables (with new columns~\citep{DasSarma:2012:FRT,Cafarella:2009:DIR,Lehmberg:2015:MSJ,Yakout:2012:IEA,Bhagavatula:2013:MEM,Zhang:2017:ESA}, rows~\citep{DasSarma:2012:FRT,Yakout:2012:IEA,Zhang:2017:ESA}, cell values~\cite{Ahmadov:2015:THI}, or links to entities~\citep{Bhagavatula:2015:TEL}).

\begin{figure}[t]
   \centering
   \includegraphics[width=0.36\textwidth]{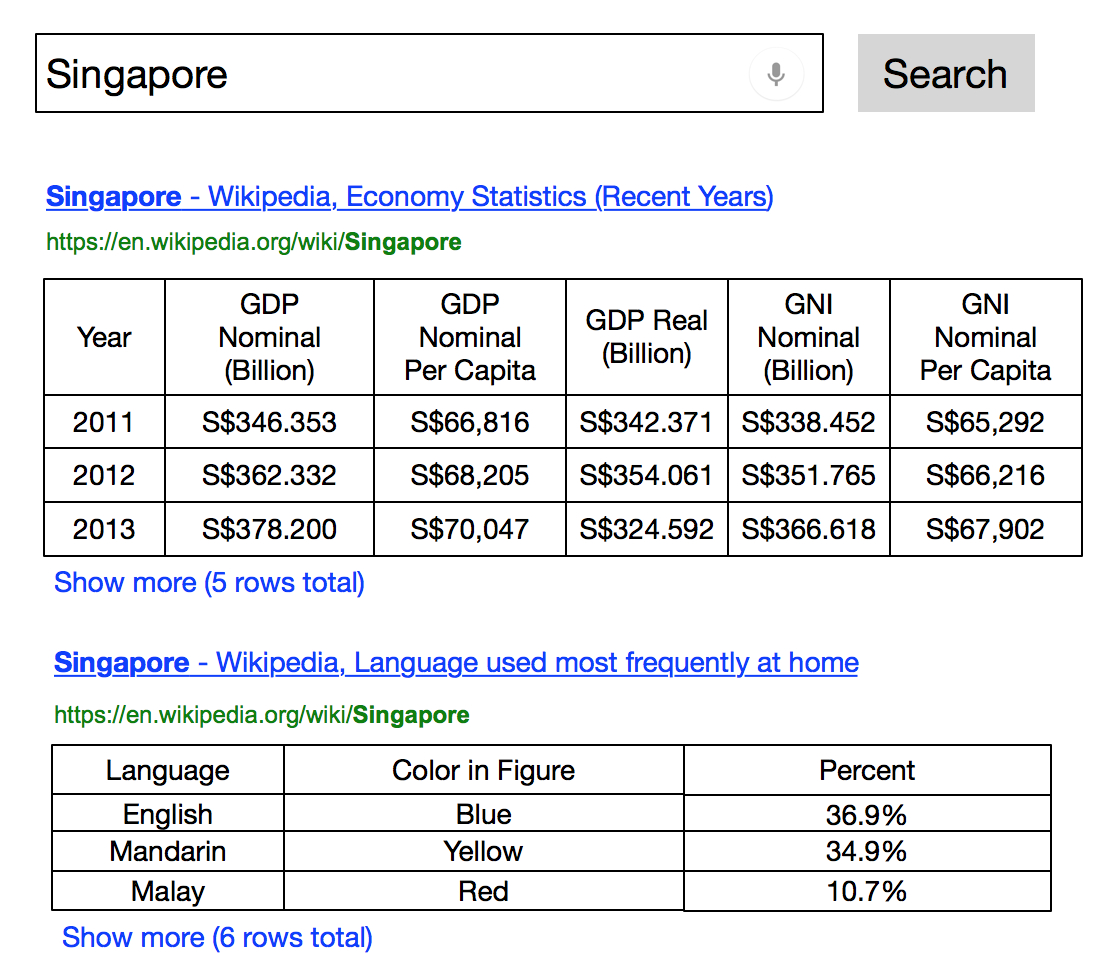} 
   \caption{Ad hoc table retrieval: given a keyword query, the system returns a ranked list of tables.}    
\label{fig:tableretrieval}
\end{figure}

Searching for tables is an important problem on its own, in addition to being a core building block in many other table-related tasks.  Yet, it has not received due attention, and especially not from an information retrieval perspective. 
This paper aims to fill that gap.  
We define the \emph{ad hoc table retrieval} task as follows: given a keyword query, return a ranked list of tables from a table corpus that are relevant to the query.
See Fig.~\ref{fig:tableretrieval} for an illustration. 

It should be acknowledged that this task is not entirely new, in fact, it has been around for a while in the database community (also known there as \emph{relation ranking})~\citep{Cafarella:2008:WEP,Cafarella:2009:DIR,Venetis:2011:RST,Bhagavatula:2013:MEM}.
However, public test collections and proper evaluation methodology are lacking, in addition to the need for better ranking techniques.

%
%
Tables can be ranked much like documents, by considering the words contained in them~\cite{Cafarella:2008:WEP, Cafarella:2009:DIR, Pimplikar:2012:ATQ}.  Ranking may be further improved by incorporating additional signals related to table quality.  Intuitively, high quality tables are topically coherent; other indicators may be related to the pages that contain them (e.g., if they are linked by other pages~\cite{Bhagavatula:2013:MEM}).
However, a major limitation of prior approaches is that they only consider lexical matching between the contents of tables and queries. 
This gives rise to our main research objective: 
\emph{Can we move beyond lexical matching and improve table retrieval performance by incorporating semantic matching?}

We consider two main kinds of semantic representations.  One is based on concepts, such as entities and categories.  Another is based on continuous vector representations of words and of entities (i.e., word and graph embeddings).  We introduce a framework that handles matching in different semantic spaces in a uniform way, by modeling both the table and the query as sets of semantic vectors.  We propose two general strategies (early and late fusion), yielding four different measures for computing the similarity between queries and tables based on their semantic representations.

As we have mentioned above, another key area where prior work has insufficiencies is evaluation.  First, there is no publicly available test collection for this task.  Second, evaluation has been performed using set-based metrics (counting the number of relevant tables in the top-$k$ results), which is a very rudimentary way of measuring retrieval effectiveness.
We address this by developing a purpose-built test collection, comprising of 1.6M tables from Wikipedia, and a set of queries with graded relevance judgments.
We establish a learning-to-rank baseline that encompasses a rich set of features from prior work, and outperforms the best approaches known in the literature.  We show that the semantic matching methods we propose can substantially and significantly improve retrieval performance over this strong baseline.

In summary, this paper makes the following contributions:

\begin{itemize}
	\item We introduce and formalize the ad hoc table ranking task, and present both unsupervised and supervised baseline approaches (Sect.~\ref{sec:tableret}). 
	\item We present a set of novel semantic matching methods that go beyond lexical similarity (Sect.~\ref{sec:sem}).
	\item We develop a standard test collection for this task (Sect.~\ref{sec:testcoll}) and demonstrate the effectiveness of our approaches (Sect.~\ref{sec:eval}).
\end{itemize}
The test collection and the outputs of the reported methods are made available at \url{https://github.com/iai-group/www2018-table}.

\section{Ad Hoc Table Retrieval}
\label{sec:tableret}

We formalize the ad hoc table retrieval task, explain what information is associated with a table, and introduce baseline methods.

\subsection{Problem Statement}

Given a keyword query $q$, \emph{ad hoc table retrieval} is the task of returning a ranked list of tables, $(T_1,\dots,T_k)$, from a collection of tables $C$.
Being an ad hoc task, the relevance of each returned table $T_i$ is assessed independently of all other returned tables $T_j, i\neq j$.
Hence, the ranking of tables boils down to the problem of assigning a score to each table in the corpus: $\mathit{score}(q,T)$.  Tables are then sorted in descending order of their scores.

\subsection{The Anatomy of a Table} \label{sec:tableret:whatis}

\begin{figure}[t]
   \centering
   \includegraphics[width=0.5\textwidth]{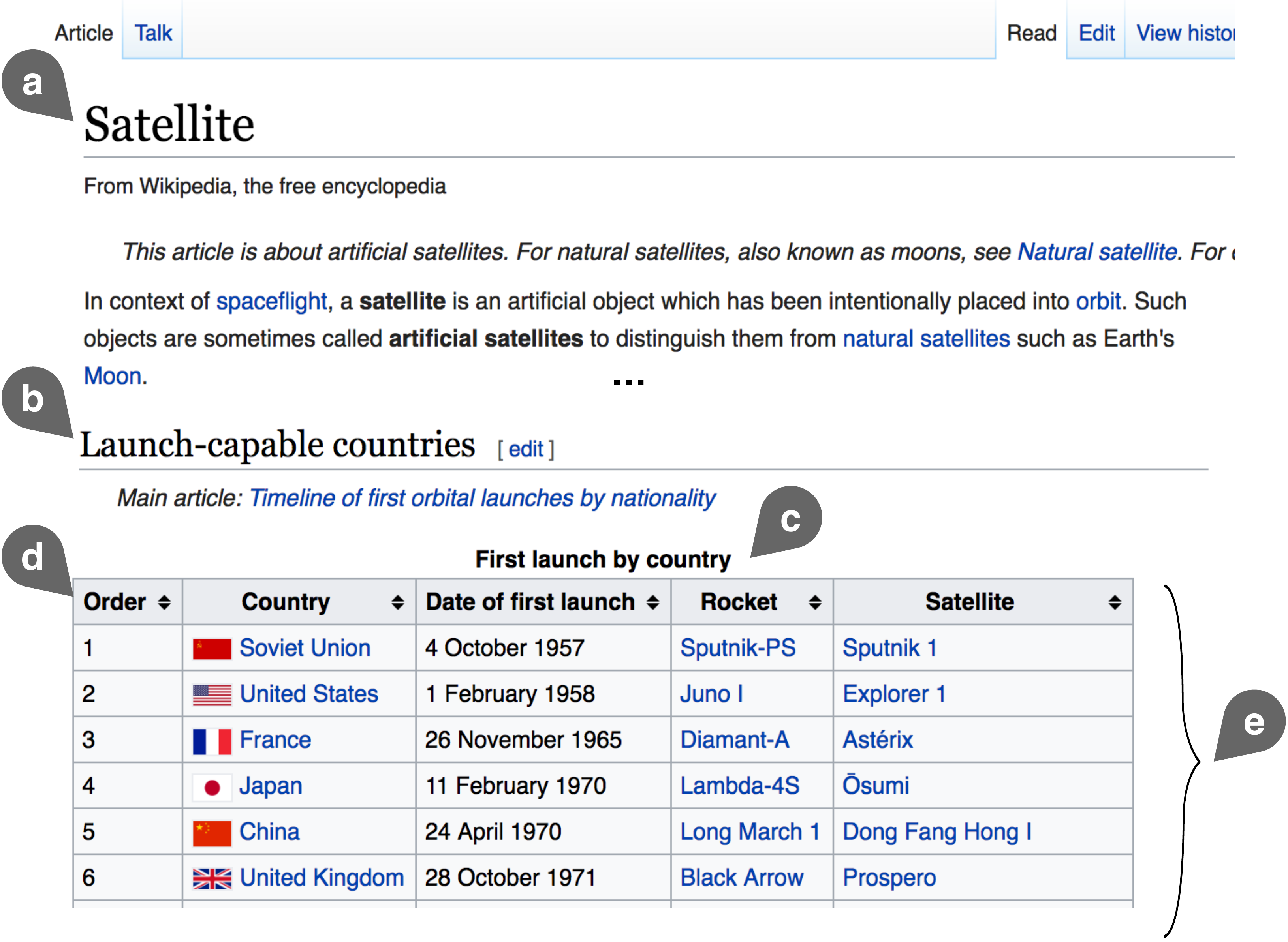} 
	\vspace*{-1.5\baselineskip}   
   \caption{Table embedded in a Wikipedia page.} 
   \label{fig:wikitable}
   \vspace*{-\baselineskip}
\end{figure}

We shall assume that the following information is available for each table in the corpus; the letters refer to Figure~\ref{fig:wikitable}.
\begin{enumerate}[label=(\alph*)]
	\item \emph{Page title}, where the table was extracted from.
	\item \emph{Section title}, i.e., the heading of the particular section where the table is embedded.
	\item \emph{Table caption}, providing a brief explanation.
	\item \emph{Table headings}, i.e., a list of column heading labels.
	\item \emph{Table body}, i.e., all table cells (including column headings).
\end{enumerate}

\subsection{Unsupervised Ranking} 
\label{sec:tbr}

An easy and straightforward way to perform the table ranking task is by adopting standard document ranking methods. 
\citet{Cafarella:2009:DIR, Cafarella:2008:WEP} utilize web search engines to retrieve relevant documents; tables are then extracted from the highest-ranked documents.  
Rather than relying on external services, we represent tables as either single- or multi-field documents and apply standard documents retrieval techniques. 

\begin{table*}[t]
\centering
\caption{Baseline features for table retrieval.}
\vspace*{-0.75\baselineskip}
\begin{tabular}{p{2.5cm}p{11.5cm}p{1cm}p{1cm}}
	\toprule
	\multicolumn{2}{l}{\textbf{Query features}} & \textbf{Source} & \textbf{Value} \\
	\midrule
	QLEN & Number of query terms & \cite{Tyree:2011:PBR} &  \{1,...,n\} \\
	$\mathrm{IDF}_f$ & Sum of query IDF scores in field $f$ & \cite{Qin:2010:LBC} & $[0,\infty)$ \\	
	\midrule
	\multicolumn{4}{l}{\textbf{Table features}} \\
	\midrule
	\#rows & The number of rows in the table & \cite{Cafarella:2008:WEP,Bhagavatula:2013:MEM} & \{1,...,n\} \\
	\#cols & The number of columns in the table & \cite{Cafarella:2008:WEP,Bhagavatula:2013:MEM} & \{1,...,n\} \\
	\#of NULLs in table & The number of empty table cells & \cite{Cafarella:2008:WEP,Bhagavatula:2013:MEM} &  \{0,...,n\} \\
	PMI & The ACSDb-based schema coherency score & \cite{Cafarella:2008:WEP} & $(-\infty,\infty)$ \\	
	inLinks & Number of in-links to the page embedding the table & \cite{Bhagavatula:2013:MEM} & \{0,...,n\} \\
	outLinks & Number of out-links from the page embedding the table & \cite{Bhagavatula:2013:MEM} & \{0,...,n\}\\
	pageViews & Number of page views & \cite{Bhagavatula:2013:MEM} & \{0,...,n\}\\
	tableImportance & Inverse of number of tables on the page & \cite{Bhagavatula:2013:MEM} & $(0,1]$ \\
	tablePageFraction & Ratio of table size to page size & \cite{Bhagavatula:2013:MEM} & $(0,1]$ \\
	\midrule
	\multicolumn{4}{l}{\textbf{Query-table features}} \\
	\midrule
	\#hitsLC & Total query term frequency in the leftmost column cells & \cite{Cafarella:2008:WEP} & \{0,...,n\} \\
	\#hitsSLC & Total query term frequency in second-to-leftmost column cells & \cite{Cafarella:2008:WEP} & \{0,...,n\} \\
	\#hitsB & Total query term frequency in the table body & \cite{Cafarella:2008:WEP} & \{0,...,n\} \\
	qInPgTitle & Ratio of the number of query tokens found in page title to total number of tokens & \cite{Bhagavatula:2013:MEM} & $[0,1]$ \\
	qInTableTitle & Ratio of the number of query tokens found in table title to total number of tokens & \cite{Bhagavatula:2013:MEM} & $[0,1]$ \\
	yRank & Rank of the table's Wikipedia page in Web search engine results for the query & \cite{Bhagavatula:2013:MEM} & \{1,...,n\} \\
	MLM similarity & Language modeling score between query and multi-field document repr. of the table & \cite{Chen:2016:ESL} & ($-\infty$,0) \\
		\bottomrule
\end{tabular}
\label{tbl:features}
\end{table*}

\subsubsection{Single-field Document Representation}
In the simplest case, all text associated with a given table is used as the table's representation.  This representation is then scored using existing retrieval methods, such as BM25 or language models. 

\subsubsection{Multi-field Document Representation}
Rather than collapsing all textual content into a single-field document, it may be organized into multiple fields, such as table caption, table headers, table body, etc. (cf. Sect.~\ref{sec:tableret:whatis}).
For multi-field ranking, \citet{Pimplikar:2012:ATQ} employ a \emph{late fusion} strategy~\cite{Shuo:2017:DPF}.  That is, each field is scored independently against the query, then a weighted sum of the field-level similarity scores is taken:

\begin{equation}
	\mathit{score}(q,T) = \sum_{i} w_i \times \mathit{score}(q, f_i) ~,
\end{equation}
where $f_i$ denotes the $i$th (document) field for table $T$ and $w_i$ is the corresponding field weight (such that $\sum_i w_i = 1$). $\mathit{score}(q, f_i)$ may be computed using any standard retrieval method.  We use language models in our experiments.

\subsection{Supervised Ranking}  
\label{sec:fbr}

The state-of-the-art in document retrieval (and in many other retrieval tasks) is to employ supervised learning~\citep{Liu:2011:LRI}.  Features may be categorized into three groups: (i) document, (ii) query, and (iii) query-document features~\citep{Qin:2010:LBC}.  Analogously, we distinguish between three types of features: (i) table, (ii) query, and (iii) query-table features.
In Table~\ref{tbl:features}, we summarize the features from previous work on table search~\citep{Cafarella:2008:WEP,Bhagavatula:2013:MEM}.  We also include a number of additional features that have been used in other retrieval tasks, such as document and entity ranking; we do not regard these as novel contributions.

\subsubsection{Query Features}
Query features have been shown to improve retrieval performance for document ranking~\citep{Macdonald:2012:UQF}.
We adopt two query features from document retrieval, namely, 
the number of terms in the query~\citep{Tyree:2011:PBR}, 
and query IDF~\citep{Qin:2010:LBC} according to:
$\mathit{IDF}_f(q) = \sum_{t \in q} \mathit{IDF}_f(t)$,
where $\mathit{IDF}_f(t)$ is the IDF score of term $t$ in field $f$.  This feature is computed for the following fields: page title, section title, table caption, table heading, table body, and ``catch-all'' (the concatenation of all textual content in the table).

\subsubsection{Table Features}
\label{subsec:tf}

Table features depend only on the table itself and aim to reflect the quality of the given table (irrespective of the query).
Some features are simple characteristics, like the number of rows, columns, and empty cells~\cite{Cafarella:2008:WEP,Bhagavatula:2013:MEM}. 
A table's PMI is computed by calculating the PMI values between all pairs of column headings of that table, and then taking their average. 
Following~\citep{Cafarella:2008:WEP}, we compute PMI by obtaining frequency statistics from the Attribute Correlation Statistics Database (ACSDb)~\cite{Cafarella:2008:URW}, which contains table heading information derived from millions of tables extracted from a large web crawl.

Another group of features has to do with the page that embeds the table, by considering its connectivity (inLinks and outLinks), popularity (pageViews), and the table's importance within the page (tableImportance and tablePageFraction).

\subsubsection{Query-Table Features}
Features in the last group express the degree of matching between the query and a given table.
This matching may be based on occurrences of query terms in the page title (qInPgTitle) or in the table caption (qInTableTitle).
Alternatively, it may be based on specific parts of the table, such as the leftmost column (\#hitsLC), second-to-left column (\#hitsSLC), or table body (\#hitsB).
Tables are typically embedded in (web) pages.  The rank at which a table's parent page is retrieved by an external search engine is also used as a feature (yRank).  (In our experiments, we use the Wikipedia search API to obtain this ranking.)
Furthermore, we take the Mixture of Language Models (MLM) similarity score~\citep{Ogilvie:2003:MLM} as a feature, which is actually the best performing method among the four text-based baseline methods (cf. Sect.~\ref{sec:eval}).
Importantly, all these features are based on lexical matching.  Our goal in this paper is to also enable semantic matching; this is what we shall discuss in the next section.

\begin{figure*}[t]
   \centering
   \includegraphics[width=0.9\textwidth]{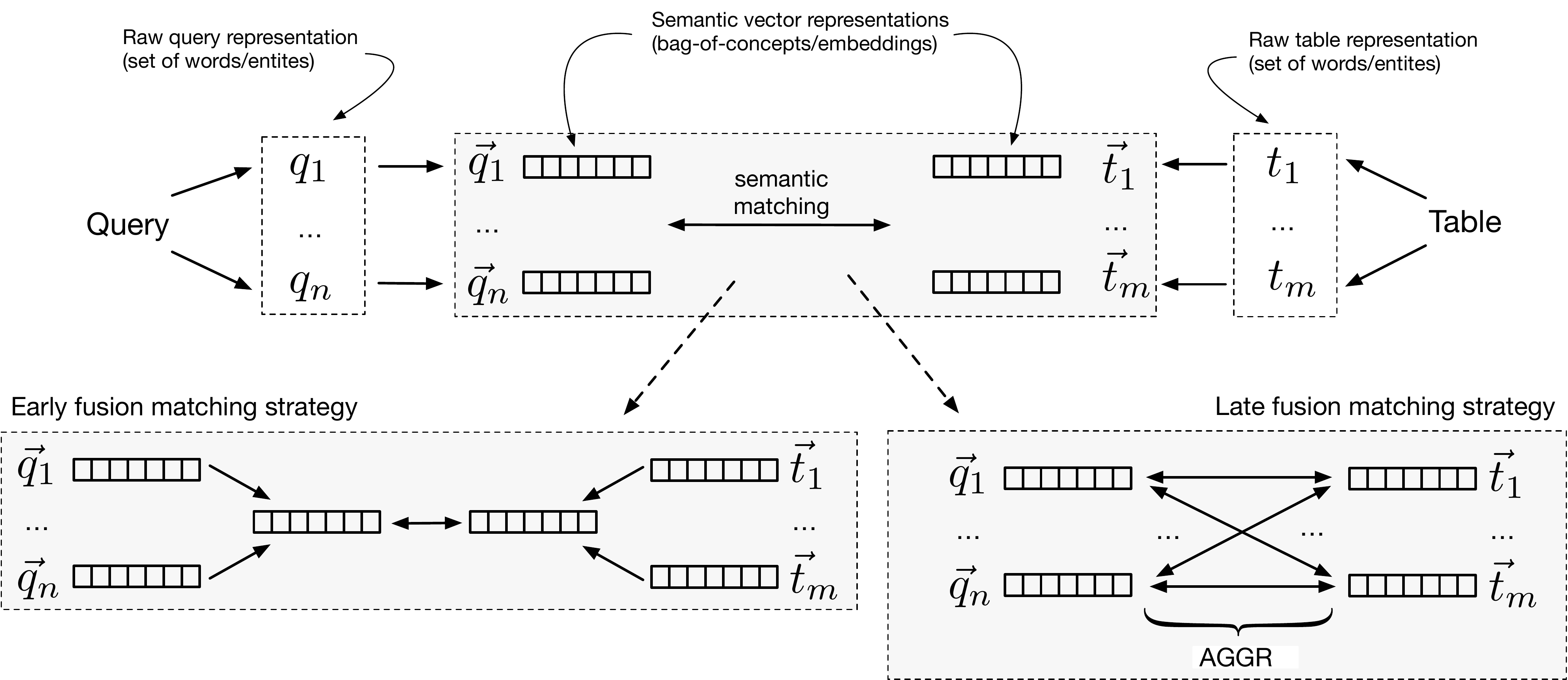} 
	\vspace*{-0.5\baselineskip}
   \caption{Our methods for computing query-table similarity using semantic representations. }
\label{fig:e3v}
\end{figure*}

\section{Semantic Matching}
\label{sec:sem}

This section presents our main contribution, which is a set of novel semantic matching methods for table retrieval.  
The main idea is to go beyond lexical matching by representing both queries and tables in some semantic space, and measuring the similarity of those semantic (vector) representations.
Our approach consists of three main steps, which are illustrated in Figure~\ref{fig:e3v}.  These are as follows (moving from outwards to inwards on the figure):
\begin{enumerate}
	\item The ``raw'' content of a query/table is represented as a set of terms, where terms can be either words or entities (Sect.~\ref{sec:sem:repr_raw}).
	\item Each of the raw terms is mapped to a semantic vector representation (Sect.~\ref{sec:sem:repr_sem}).
	\item The semantic similarity (matching score) between a query-table pair is computed based on their semantic vector representations (Sect.~\ref{sec:sem:match}).
\end{enumerate}
We compute query-table similarity using all possible combinations of semantic representations and similarity measures, and use the resulting semantic similarity scores as features in a learning-to-rank approach.  Table~\ref{tbl:features2} summarizes these features.

%
%

\subsection{Content Extraction} 
\label{sec:sem:repr_raw}

We represent the ``raw'' content of the query/table as a set of terms, where terms can be either words (string tokens) or entities (from a  knowledge base).  We denote these as $\{q_1, \dots, q_n\}$ and $\{t_1, \dots, t_m\}$ for query $q$ and table $T$, respectively.  

\subsubsection{Word-based}
It is a natural choice to simply use word tokens to represent query/table content.  That is, $\{q_1, \dots, q_n\}$ is comprised of the unique words in the query. 
As for the table, we let $\{t_1, \dots, t_m\}$ contain all unique words from the title, caption, and headings of the table.
Mind that at this stage we are only considering the presence/absence of words.  During the query-table similarity matching, the importance of the words will also be taken into account (Sect.~\ref{sec:sem:match_early}).

\subsubsection{Entity-based}
Many tables are focused on specific entities~\cite{Zhang:2017:ESA}.  Therefore, considering the entities contained in a table amounts to a meaningful   representation of its content.
We use the DBpedia knowledge base as our entity repository. 
Since we work with tables extracted from Wikipedia, the entity annotations are readily available (otherwise, entity annotations could be obtained automatically, see, e.g., ~\cite{Venetis:2011:RST}).
Importantly, instead of blindly including all entities mentioned in the table, we wish to focus on salient entities.
It has been observed in prior work~\cite{Venetis:2011:RST, Bhagavatula:2015:TEL} that tables often have a \emph{core column}, containing mostly entities, while the rest of the columns contain properties of these entities (many of which are entities themselves).
We write $E_{cc}$ to denote the set of entities that are contained in the core column of the table, and describe our core column detection method in Sect.~\ref{sec:sem:core_col}. 
In addition to the entities taken directly from the body part of the table, we also include entities that are related to the page title ($T_{pt}$) and to the table caption ($T_{tc}$).  We obtain those by using the page title and the table caption, respectively, to retrieve relevant entities from the knowledge base.  We write $R_k(s)$ to denote the set of top-$k$ entities retrieved for the query $s$.  We detail the entity ranking method in Sect.~\ref{sec:sem:er}.  
Finally, the table is represented as the union of three sets of entities, originating from the core column, page title, and table caption: 
$\{t_1, \dots, t_m\} = E_{cc} \cup R_k(T_{pt}) \cup R_k(T_{tc})$.

To get an entity-based representation for the query, we issue the query against a knowledge base to retrieve relevant entities, using the same retrieval method as above.  I.e., $\{q_1, \dots, q_n\} = R_k(q)$.

\subsubsection{Core Column Detection}
\label{sec:sem:core_col}

We introduce a simple and effective core column detection method.  It is based on the notion of \emph{column entity rate}, which is defined as the ratio of cells in a column that contain an entity.  We write $\mathit{cer}(T_{c[j]})$ to denote the column entity rate of column $j$ in table $T$.
Then, the index of the core column becomes: 
$\arg\max_{j=1..T_{|c|}} \mathit{cer}(T_{c[j]})$,
where $T_{|c|}$ is the number of columns in $T$.

\subsubsection{Entity Retrieval}
\label{sec:sem:er}

We employ a fielded entity representation with five fields (names, categories, attributes, similar entity names, and related entity names) and rank entities using the Mixture of Language Models approach~\cite{Ogilvie:2003:MLM}.  The field weights are set uniformly. This corresponds to the MLM-all model in~\citep{Hasibi:2017:DVT} and is shown to be a solid baseline.
We return the top-$k$ entities, where $k$ is set to 10.

\begin{table}[t]
\centering
\caption{Semantic similarity features.
Each row represents 4 features (one for each similarity matching method, cf. Table~\ref{tbl:sim_measures}).  All features are in $[-1,1]$.}
\vspace*{-0.5\baselineskip}
\begin{tabular}{p{1.5cm}p{2.5cm}p{1.4cm}}
	\toprule
	\textbf{Features} & \textbf{Semantic repr.} & \textbf{Raw repr.} \\
	\midrule
		Entity\_* & Bag-of-entities & entities \\ 
		Category\_* & Bag-of-categories & entities \\ 
		Word\_* & Word embeddings & words \\ 
		Graph\_* & Graph embeddings & entities \\ 
	\bottomrule
\end{tabular}
\vspace*{-0.5\baselineskip}
\label{tbl:features2}
\end{table}

\subsection{Semantic Representations}
\label{sec:sem:repr_sem}

Next, we embed the query/table terms in a semantic space. That is, we map each table term $t_i$ to a vector representation $\vec{t}_i$, where $\vec{t}_i[j]$ refers to the $j$th element of that vector.  For queries, the process goes analogously. 
We discuss two main kinds of semantic spaces, bag-of-concepts and embeddings, with two alternatives within each.  The former uses sparse and discrete, while the latter employs dense and continuous-valued vectors.  A particularly nice property of our semantic matching framework is that it allows us to deal with these two different types of representations in a unified way.

\subsubsection{Bag-of-concepts}
One alternative for moving from the lexical to the semantic space is to represent tables/queries using specific concepts. In this work, we use entities and categories from a knowledge base.
These two semantic spaces have been used in the past for various retrieval tasks, in duet with the traditional bag-of-words content representation.
For example, entity-based representations have been used for document retrieval~\citep{Xiong:2017:WDR,Raviv:2016:DRU} and category-based representations have been used for entity retrieval~\citep{Balog:2011:QME}.
One important difference from previous work is that instead of representing the entire query/table using a single semantic vector, we map each individual query/table term to a separate semantic vector, thereby obtaining a richer representation.  

We use the entity-based raw representation from the previous section, that is, $t_i$ and $q_j$ are specific entities. Below, we explain how table terms $t_j$ are projected to $\vec{t_i}$, which is a sparse discrete vector in the entity/category space; for query terms it follows analogously.

\begin{description}
	\item[Bag-of-entities] Each element in $\vec{t}_i$ corresponds to a unique entity. Thus, the dimensionality of $\vec{t}_i$ is the number of entities in the knowledge base (on the order of millions). $\vec{t}_i[j]$ has a value of $1$ if entities $i$ and $j$ are related (there exists a link between them in the knowledge base), and $0$ otherwise.  
	\item[Bag-of-categories] Each element in $\vec{t}_i$ corresponds to a Wikipedia category.  Thus, the dimensionality of $\vec{t}_i$ amounts to the number of Wikipedia categories (on the order hundreds of thousands).  The value of $\vec{t}_i[j]$ is $1$ if entity $i$ is assigned to Wiki\-pe\-dia category $j$, and $0$ otherwise.
\end{description}

\subsubsection{Embeddings}
\label{sec:sem:embeddings}

Recently, unsupervised representation learning methods have been proposed for obtaining embeddings that predict a distributional context, i.e., word embeddings~\citep{Mikolov:2013:DRW,Pennington:2014:GGV} or graph embeddings~\citep{Perozzi:2014:DOL,Tang:2015:LLI,Ristoski:2016:RGE}.  Such vector representations have been utilized successfully in a range of IR tasks, including ad hoc retrieval~\cite{Ganguly:2015:WEB, Bhaskar:2016:DES}, contextual suggestion~\cite{Jarana:2016:MUP}, cross-lingual IR~\cite{Vulic:2015:MCI}, community question answering~\cite{zhou-EtAl:2015:ACL-IJCNLP1}, short text similarity~\cite{Kenter:2015:STS}, and sponsored search~\cite{Grbovic:2015:CCE}.
We consider both word-based and entity-based raw representations from the previous section and use the corresponding (pre-trained) embeddings as follows.

\begin{description}
	\item[Word embeddings] We map each query/table word to a word embedding. Specifically, we use word2vec~\citep{Mikolov:2013:DRW} with 300 dimensions, trained on Google News data. 
	\item[Graph embeddings] We map each query/table entity to a graph embedding.  In particular, we use RDF2vec~\cite{Ristoski:2016:RGE} with 200 dimensions, trained on DBpedia 2015-10.
\end{description}

\subsection{Similarity Measures}
\label{sec:sem:match}

The final step is concerned with the computation of the similarity between a query-table pair, based on the semantic vector representations we have obtained for them.  We introduce two main strategies, which yield four specific similarity measures. These are summarized in Table~\ref{tbl:sim_measures}.

\subsubsection{Early Fusion}
\label{sec:sem:match_early}

The first idea is to represent the query and the table each with a single vector.  Their similarity can then simply be expressed as the similarity of the corresponding vectors.  We let $\vec{C}_q$ be the centroid of the query term vectors ($\vec{C}_q = \sum_{i=1}^n \vec{q}_i/n$).
Similarly, $\vec{C}_T$ denotes the centroid of the table term vectors.
The query-table similarity is then computed by taking the cosine similarity of the centroid vectors.
%
%
When query/table content is represented in terms of words, we additionally make use of word importance by employing standard TF-IDF term weighting.  Note that this only applies to word embeddings (as the other three semantic representations are based on entities).  In case of word embeddings, the centroid vectors are calculated as $\vec{C}_T = \sum_{i=1}^m \vec{t}_i \times TFIDF (t_i)$. The computation of $\vec{C}_q$ follows analogously.

\begin{table}[t]
\centering
\caption{Similarity measures.}
\vspace*{-0.75\baselineskip}
\begin{tabular}{p{1.5cm}p{5.5cm}}
	\toprule
	\textbf{Measure}  & \textbf{Equation} \\
	\midrule
	Early  & $\cos(\vec{C}_{q}, \vec{C}_{T})$ \\
	Late-max & $\mathrm{max}(\{\cos(\vec{q}_i,\vec{t}_j) : i \in [1..n], j \in [1..m] \})$\\
	Late-sum & $\mathrm{sum}(\{\cos(\vec{q}_i,\vec{t}_j) : i \in [1..n], j \in [1..m] \})$\\
	Late-avg & $\mathrm{avg}(\{\cos(\vec{q}_i,\vec{t}_j) : i \in [1..n], j \in [1..m] \})$\\
	\bottomrule
\end{tabular}
\label{tbl:sim_measures}
\end{table}

\subsubsection{Late Fusion}
\label{sec:sem:match_late}

Instead of combining all semantic vectors $q_i$ and $t_j$ into a single one, late fusion computes the pairwise similarity between all query and table  vectors first, and then aggregates those.  We let $S$ be a set that holds all pairwise cosine similarity scores: 
$S = \{\cos(\vec{q}_i,\vec{t}_j) : i \in [1..n], j \in [1..m] \}$.
The query-table similarity score is then computed as $\mathrm{aggr}(S)$, where $\mathrm{aggr}()$ is an aggregation function.  Specifically, we use $\max()$, $\mathrm{sum}()$ and $\mathrm{avg}()$ as aggregators; see the last three rows in Table~\ref{tbl:sim_measures} for the equations. 

\section{Test Collection}
\label{sec:testcoll}

We introduce our test collection, including the table corpus, test and development query sets, and the procedure used for obtaining relevance assessments.

\subsection{Table Corpus}
\label{sec:testcoll:tablecorpus}

We use the WikiTables corpus~\cite{Bhagavatula:2015:TEL}, which comprises 1.6M tables extracted from Wikipedia (dump date: 2015 October). 
The following information is provided for each table: table caption, column headings, table body, (Wikipedia) page title, section title, and table statistics like number of headings rows, columns, and data rows.
We further replace all links in the table body with entity identifiers from the DBpedia knowledge base (version 2015-10) as follows.  For each cell that contains a hyperlink, we check if it points to an entity that is present in DBpedia.  If yes, we use the DBpedia identifier of the linked entity as the cell's content; otherwise, we replace the link with the anchor text, i.e., treat it as a string.


\subsection{Queries}
\label{subsec:testcoll:queries}



We sample a total of 60 test queries from two independent sources (30 from each):
(1) \emph{Query subset 1 (QS-1)}: \citet{Cafarella:2009:DIR} collected 51 queries from Web users via crowdsourcing (using Amazon's Mechanical Turk platform, users were asked to suggest topics or supply URLs for a useful data table). 
(2) \emph{Query subset 2 (QS-2)}: \citet{Venetis:2011:RST} analyzed the query logs from Google Squared (a service in which users search for structured data) and constructed 100 queries, all of which are a combination of an instance class (e.g., ``laptops'') and a property (e.g., ``cpu'').
Following~\citep{Bhagavatula:2013:MEM}, we concatenate the class and property fields into a single query string (e.g., ``laptops cpu''). 
Table~\ref{tbl:queries} lists some examples. 

\begin{table}[t]
\centering
\caption{Example queries from our query set.}
\vspace*{-0.75\baselineskip}
\begin{tabular}{p{3cm}p{3cm}}
	\toprule
	\textbf{Queries from~\cite{Cafarella:2009:DIR}}  & \textbf{Queries from~\cite{Venetis:2011:RST}} \\
	\midrule
	video games  & asian coutries currency \\
	us cities & laptops cpu\\
	kings of africa & food calories\\
	economy gdp & guitars manufacturer\\
	fifa world cup winners&clothes brand\\	
	\bottomrule
\end{tabular}
\label{tbl:queries}
\end{table}
\label{sec:data:query}

\subsection{Relevance Assessments}
\label{subsec:testcoll:rel}


We collect graded relevance assessments by employing three independent (trained) judges.  
For each query, we pool the top 20 results from five baseline methods (cf. Sect.~\ref{sec:eval:baselines}), using default parameter settings. (Then, we train the parameters of those methods with help of the obtained relevance labels.)
Each query-table pair is judged on a three point scale: 0 (non-relevant), 1 (somewhat relevant), and 2 (highly relevant).
Annotators were situated in a scenario where they need to create a table on the topic of the query, and wish to find relevant tables that can aid them in completing that task.
Specifically, they were given the following labeling guidelines:
(i) a table is \emph{non-relevant} if it is unclear what it is about (e.g., misses headings or caption) or is about a different topic; 
(ii) a table is \emph{relevant} if some cells or values could be used from this table; and 
(iii) a table is \emph{highly relevant} if large blocks or several values could be used from it when creating a new table on the query topic. 

We take the majority vote as the relevance label; if no majority agreement is achieved, 
we take the average of the scores as the final label.
To measure inter-annotator agreement, we compute the Kappa test statistics on test annotations, which is 0.47. According to~\citep{Fleiss:1971:MNS}, this is considered as moderate agreement. 
In total, 3120 query-table pairs are annotated as test data. 
Out of these, 377 are labeled as highly relevant, 474 as relevant, and 2269 as non-relevant.


%
%


\section{Evaluation}
\label{sec:eval}

\begin{table*}[t]
  \centering
  \caption{Table retrieval evaluation results.}
  \vspace*{-0.75\baselineskip}
  \begin{tabular}{ l  llll }
    \toprule 
	\textbf{Method} & \textbf{NDCG@5} & \textbf{NDCG@10} & \textbf{NDCG@15} & \textbf{NDCG@20} \\
	\midrule
	Single-field document ranking& 0.4315  & 0.4344  & 0.4586  & 0.5254  \\
	Multi-field document ranking & 0.4770 & 0.4860 & 0.5170 & 0.5473 \\
	WebTable~\citep{Cafarella:2008:WEP} & 0.2831  & 0.2992  & 0.3311  & 0.3726 \\
	WikiTable~\citep{Bhagavatula:2013:MEM} & 0.4903 & 0.4766 & 0.5062 & 0.5206 \\
	LTR baseline (this paper) 
		& 0.5527 & 0.5456 & 0.5738 & 0.6031 \\
	\midrule
	STR (this paper) 
		& \textbf{0.5951} & \textbf{0.6293}$^\dag$ & \textbf{0.6590}$^\ddag$ & \textbf{0.6825}$^\dag$ \\
	
    \bottomrule
  \end{tabular}
  \label{tbl:results}
\end{table*}

In this section, we list our research questions (Sect.~\ref{sec:eval:rq}), discuss our experimental setup (Sect.~\ref{sec:eval:setup}), introduce the baselines we compare against (Sect.~\ref{sec:eval:baselines}), and present our results (Sect.~\ref{sec:eval:results}) followed by further analysis (Sect.~\ref{sec:analysis}).

\subsection{Research Questions}
\label{sec:eval:rq}

The research questions we seek to answer are as follows.

\begin{description}
	\item[RQ1] Can semantic matching improve retrieval performance?
	\item[RQ2] Which of the semantic representations is the most effective?
	\item[RQ3] Which of the similarity measures performs better? 
\end{description}

\subsection{Experimental Setup} 
\label{sec:eval:setup}

We evaluate table retrieval performance in terms of Normalized Discounted Cumulative Gain (NDCG) at cut-off points 5, 10, 15, and 20.  To test significance, we use a two-tailed paired t-test and write $\dag$/$\ddag$ to denote significance at the 0.05 and 0.005 levels, respectively.

Our implementations are based on Nordlys~\citep{Hasibi:2017:NTE}. Many of our features involve external sources, which we explain below. 
To compute the entity-related features (i.e., features in Table~\ref{tbl:features} as well as the features based on the bag-of-entities and bag-of-categories representations in Table~\ref{tbl:features2}), we use entities from the DBpedia knowledge base that have an abstract (4.6M in total). 
The table's Wikipedia rank (yRank) is obtained using Wikipedia's MediaWiki API.
The PMI feature is estimated based on the ACSDb corpus~\cite{Cafarella:2008:URW}.
For the distributed representations, we take pre-trained embedding vectors, as explained in Sect.~\ref{sec:sem:embeddings}.

%
\begin{table*}[t]
  \centering
  \caption{Comparison of semantic features, used in combination with baseline features (from Table~\ref{tbl:features}), in terms of NDCG@20.  Relative improvements are shown in parentheses.
  Statistical significance is tested against the LTR baseline in Table~\ref{tbl:results}.}
  \vspace*{-0.75\baselineskip}
  \begin{tabular}{ l lllll }
    \toprule 
	\textbf{Sem. Repr.} 
	& \textbf{Early} 
	& \textbf{Late-max} 
	& \textbf{Late-sum} 
	& \textbf{Late-avg} 
	& \textbf{ALL} \\
	\midrule
	Bag-of-entities 
		& 0.6754 (+11.99\%) 
		& 0.6407 (+6.23\%)$^\dag$ 
		& 0.6697 (+11.04\%)$^\ddag$ 
		& 0.6733 (+11.64\%)$^\ddag$ 
		& \cellcolor[gray]{0.85}0.6696 (+11.03\%)$^\ddag$ \\
	Bag-of-categories 
		& 0.6287 (+4.19\%) 
		& 0.6245 (+3.55\%)  
		& 0.6315 (+4.71\%)$^\dag$ 
		& 0.6240 (+3.47\%)   
		& \cellcolor[gray]{0.85}0.6149 (+1.96\%) \\
	Word embeddings 
		& 0.6181 (+2.49\%)  
		& 0.6328 (+4.92\%)  
		& 0.6371 (+5.64\%)$^\dag$ 
		& 0.6485 (+7.53\%)$^\dag$ 
		& \cellcolor[gray]{0.85}0.6588 (+9.24\%)$^\dag$ \\
	Graph embeddings 
		& 0.6326 (+4.89\%)  
		& 0.6142 (+1.84\%)  
		& 0.6223 (+3.18\%)
		& 0.6316 (+4.73\%)  
		& \cellcolor[gray]{0.85}0.6340 (+5.12\%) \\
	ALL 
		& \cellcolor[gray]{0.85}0.6736 (+11.69\%)$^\dag$ 
		& \cellcolor[gray]{0.85}0.6631 (+9.95\%)$^\dag$ 
		& \cellcolor[gray]{0.85}0.6831 (+13.26\%)$^\ddag$ 
		& \cellcolor[gray]{0.85}0.6809 (+12.90\%)$^\ddag$  
		& \cellcolor[gray]{0.85}0.6825 (13.17\%)$^\ddag$ \\
	\bottomrule
  \end{tabular}
  \label{tbl:results2}
\end{table*}

\subsection{Baselines}
\label{sec:eval:baselines}

We implement four baseline methods from the literature.

\begin{description}
	\item[Single-field document ranking] In \cite{Cafarella:2009:DIR, Cafarella:2008:WEP} tables are represented and ranked as ordinary documents. Specifically, we use Language Models with Dirichlet smoothing, and optimize the smoothing parameter using a parameter sweep.

	\item[Multi-field document ranking] \citet{Pimplikar:2012:ATQ} represent each table as a fielded document, using five fields: Wikipedia page title, table section title, table caption, table body, and table headings.  We use the Mixture of Language Models approach~\cite{Ogilvie:2003:MLM} for ranking.  Field weights are optimized using the coordinate ascent algorithm; smoothing parameters are trained for each field individually. 

%

	\item[WebTable] 
The method by \citet{Cafarella:2008:WEP} uses the features in Table~\ref{tbl:features} with \citep{Cafarella:2008:WEP} as source.  Following~\citep{Cafarella:2008:WEP}, we train a linear regression model with 5-fold cross-validation.

	\item[WikiTable] The approach by \citet{Bhagavatula:2013:MEM} uses the features in Table~\ref{tbl:features} with \cite{Bhagavatula:2013:MEM} as source.  We train a Lasso model with coordinate ascent with 5-fold cross-validation. 

\end{description}
Additionally, we introduce a learning-to-rank baseline:
\begin{description}
	\item[LTR baseline] It uses the full set of features listed in Table~\ref{tbl:features}.  We employ pointwise regression using the Random Forest algorithm.\footnote{We also experimented with  Gradient Boosting regression and Support Vector Regression, and observed the same general patterns regarding feature importance. However, their overall performance was lower than that of Random Forests.}
We set the number of trees to 1000 and the maximum number of features in each tree to 3.  We train the model using 5-fold cross-validation (w.r.t. NDCG@20); reported results are averaged over 5 runs.
\end{description}
The baseline results are presented in the top block of Table~\ref{tbl:results}. 
It can be seen from this table that our LTR baseline (row five) outperforms all existing methods from the literature; the differences are substantial and statistically significant. 
Therefore, in the remainder of this paper, we shall compare against this strong baseline, using the same learning algorithm (Random Forests) and parameter settings.
We note that our emphasis is on the semantic matching features and not on the supervised learning algorithm.


\subsection{Experimental Results}
\label{sec:eval:results}

The last line of Table~\ref{tbl:results} shows the results for our semantic table retrieval (STR) method.  It combines the baseline set of features (Table~\ref{tbl:features}) with the set of novel semantic matching features (from Table~\ref{tbl:features2}, 16 in total).  
We find that these semantic features bring in substantial and statistically significant improvements over the LTR baseline.  Thus, we answer RQ1 positively. The relative improvements range from 7.6\% to 15.3\%, depending on the rank cut-off.

To answer RQ2 and RQ3, we report on all combinations of semantic representations and similarity measures in Table~\ref{tbl:results2}. In the interest of space, we only report on NDCG@20; the same trends were observed for other NDCG cut-offs.  Cells with a white background show retrieval performance when extending the LTR baseline with a single feature.  Cells with a grey background correspond to using a given semantic representation with different similarity measures (rows) or using a given similarity measure with different semantic representations (columns).  
The first observation is that all features improve over the baseline, albeit not all of these improvements are statistically significant.
Concerning the comparison of different semantic representations (RQ2), we find that bag-of-entities and word embeddings achieve significant improvements; see the rightmost column of Table~\ref{tbl:results2}.  
It is worth pointing out that for word embeddings the four similarity measures seem to complement each other, as their combined performance is better than that of any individual method.  It is not the case for bag-of-entities, where only one of the similarity measures (Late-max) is improved by the combination.
Overall, in answer to RQ2, we find the bag-of-entities representation to be the most effective one.  The fact that this sparse representation outperforms word embeddings is regarded as a somewhat surprising finding, given that the latter has been trained on massive amounts of (external) data.


As for the choice of similarity measure (RQ3), it is difficult to name a clear winner when a single semantic representation is used. The relative differences between similarity measures are generally small (below 5\%).
When all four semantic representations are used (bottom row in Table~\ref{tbl:results2}), we find that Late-sum and Late-avg achieve the highest overall improvement. 
Importantly, when using all semantic representations, all four similarity measures improve significantly and substantially over the baseline.  
We further note that the combination of all similarity measures do not yield further improvements over Late-sum or Late-avg.
In answer to RQ3, we identify the late fusion strategy with sum or avg aggregation (i.e., Late-sum or Late-avg) as the preferred similarity method.
\subsection{Analysis}
\label{sec:analysis}

We continue with further analysis of our results. 

\subsubsection{Features}
\label{sec:fa}
\begin{figure*}[!t]
   \centering
   \includegraphics[width=0.8\textwidth]{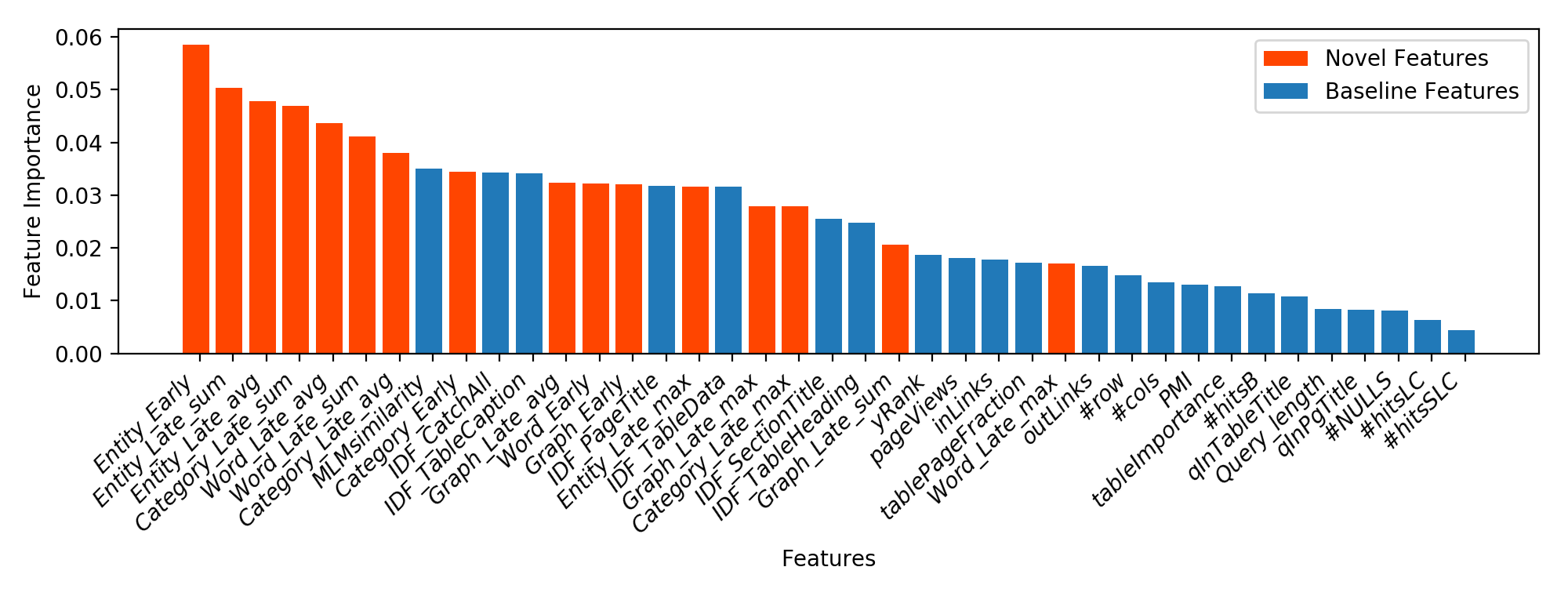} 
   \vspace*{-\baselineskip}
   \caption{Normalized feature importance (measured in terms of Gini score).}
   \vspace*{-0.5\baselineskip}
\label{fig:f_imp}
\end{figure*}

Figure~\ref{fig:f_imp} shows the importance of individual features for the table retrieval task, measured in terms of Gini importance. The novel features are distinguished by color.
We observe that 8 out of the top 10 features are semantic features introduced in this paper.  





\begin{figure*}[t]
	\centering
	\begin{tabular}{cccc}
		\subfigure[Bag-of-entities]{\includegraphics[width=0.23\textwidth]{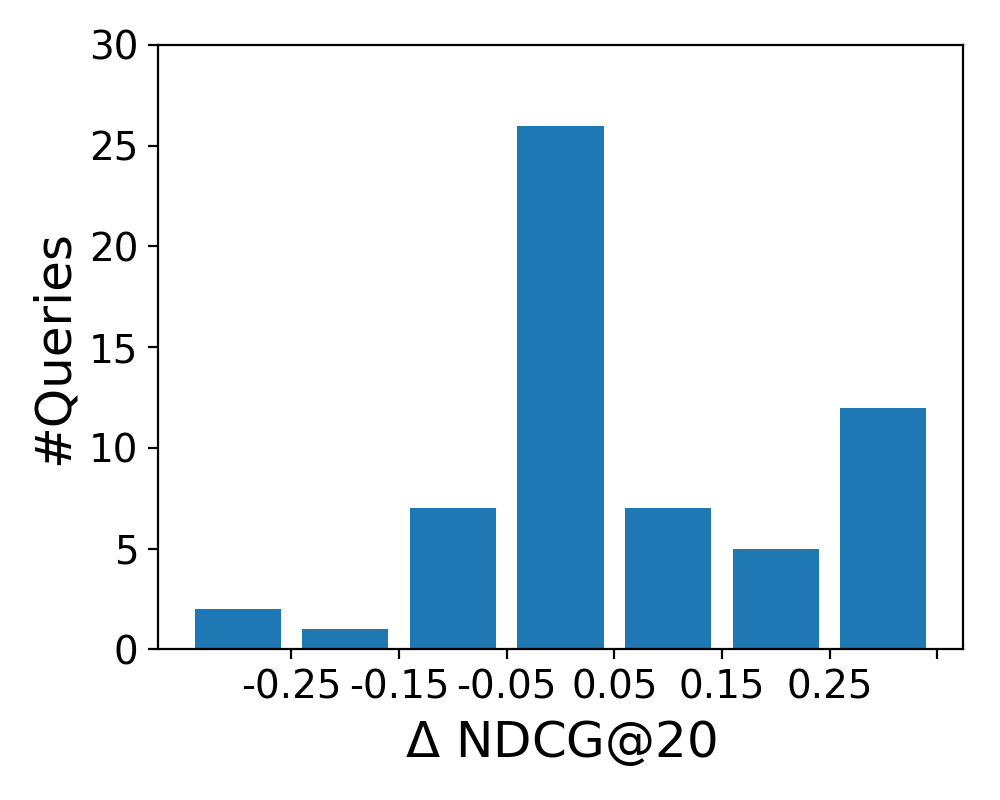}}
		&
		\subfigure[Bag-of-categories]{\includegraphics[width=0.23\textwidth]{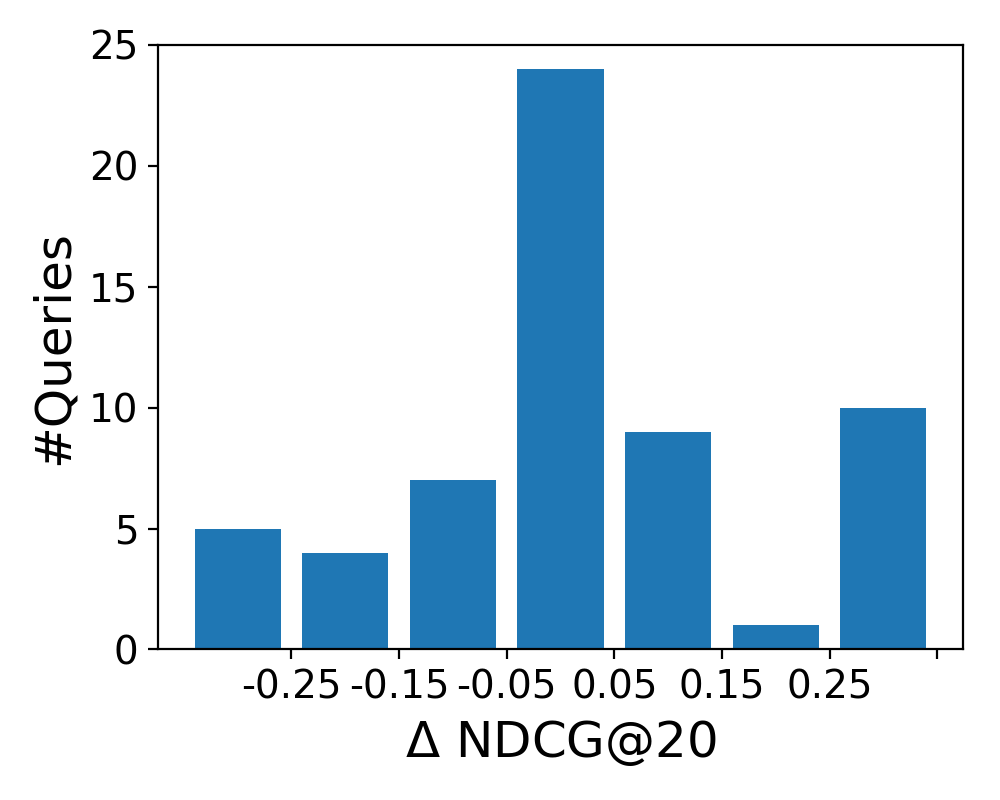}}
		&
		\subfigure[Word embeddings]{\includegraphics[width=0.23\textwidth]{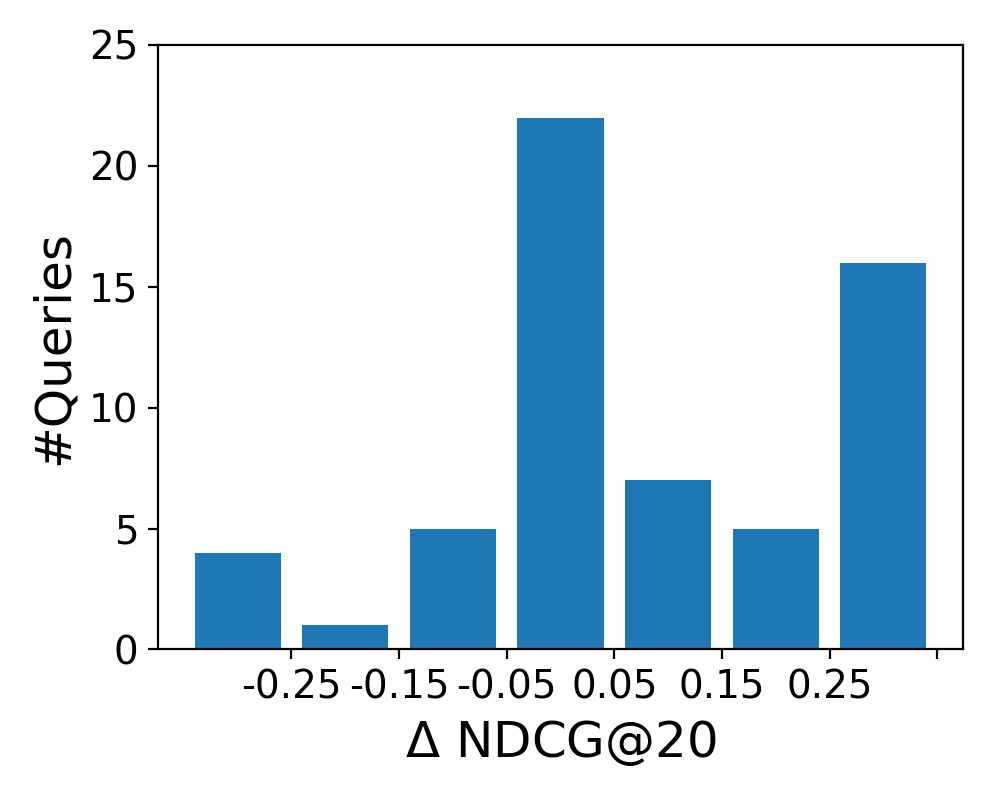}}
		&
		\subfigure[Graph embeddings]{\includegraphics[width=0.23\textwidth]{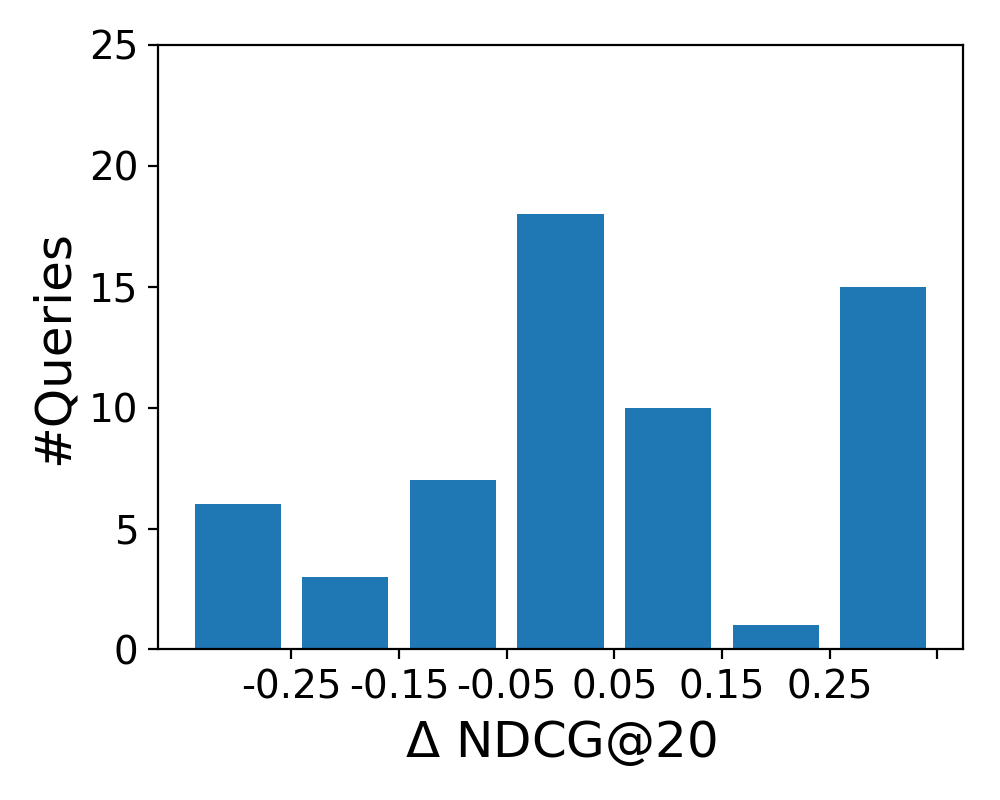}}
	\end{tabular}
	\vspace*{-\baselineskip}
	\caption{Distribution of query-level differences between the LTR baseline and a given semantic representation.}
	 \label{fig:ndcg_diff2}
\end{figure*}

\subsubsection{Semantic Representations}

To analyze how the four semantic representations affect retrieval performance on the level of individual queries, we plot the difference between the LTR baseline and each semantic representation in Figure~\ref{fig:ndcg_diff2}.  The histograms show the distribution of queries according to NDCG@20 score difference ($\Delta$): the middle bar represents no change ($\Delta<$0.05), while the leftmost and rightmost bars represents the number of queries that were hurt and helped substantially, respectively ($\Delta>$0.25).
We observe similar patterns for the bag-of-entities and word embeddings representations; the former has less queries that were significantly helped or hurt, while the overall improvement (over all topics) is larger.
We further note the similarity of the shapes of the distributions for bag-of-categories and graph embeddings.

\subsubsection{Query Subsets}
\begin{figure}[t]
   \centering
   \includegraphics[width=0.3\textwidth]{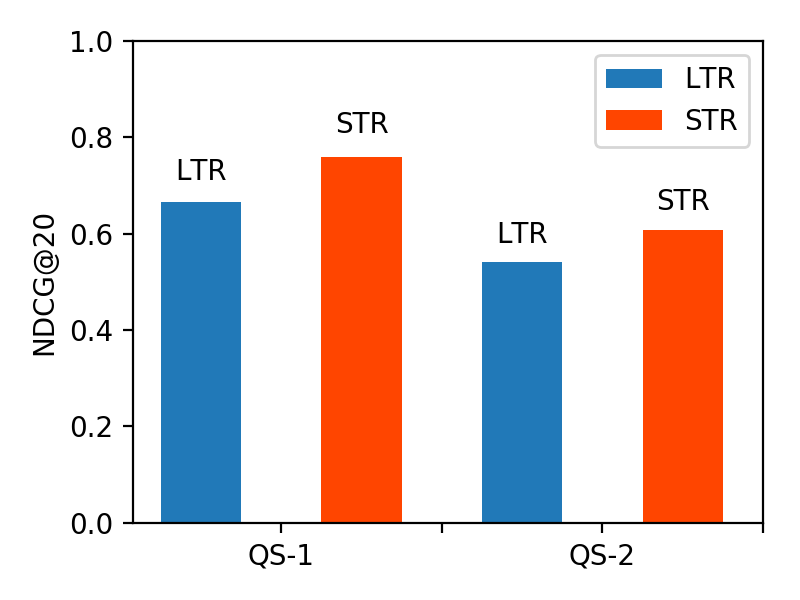} 
   \vspace*{-\baselineskip}   
   \caption{Table retrieval results, LTR baseline vs. STR, on the two query subsets in terms of NDCG@20.}
\label{fig:q1_VS_q2}
\end{figure}

On Figure~\ref{fig:q1_VS_q2}, we plot the results for the LTR baseline and for our STR method according to the two query subsets, QS-1 and QS-2, in terms of NDCG@20.  Generally, both methods perform better on QS-1 than on QS-2. This is mainly because QS-2 queries are more focused (each targeting a specific type of instance, with a required property), and thus are considered more difficult. Importantly, STR achieves consistent improvements over LTR on both query subsets.

%

%

%
\subsubsection{Individual Queries}

\begin{figure*}[t]
	\centering
	\begin{tabular}{cc}
		\subfigure[QS-1]{\includegraphics[width=0.45\textwidth]{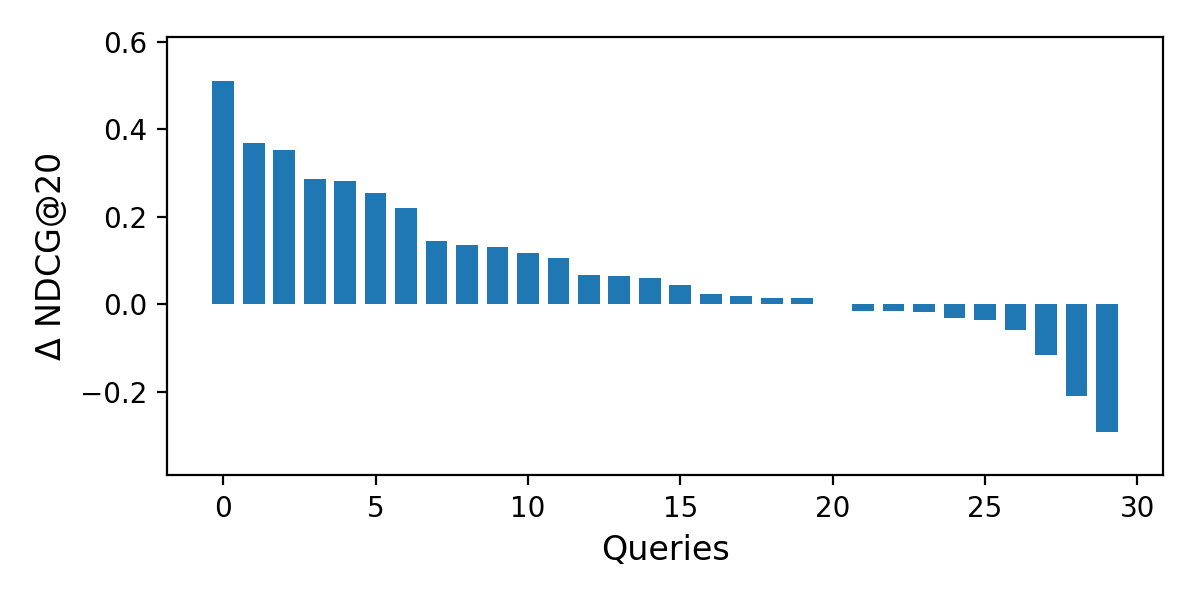}\label{fig:ndcg_diff_q1}}
		&
		\subfigure[QS-2]{\includegraphics[width=0.45\textwidth]{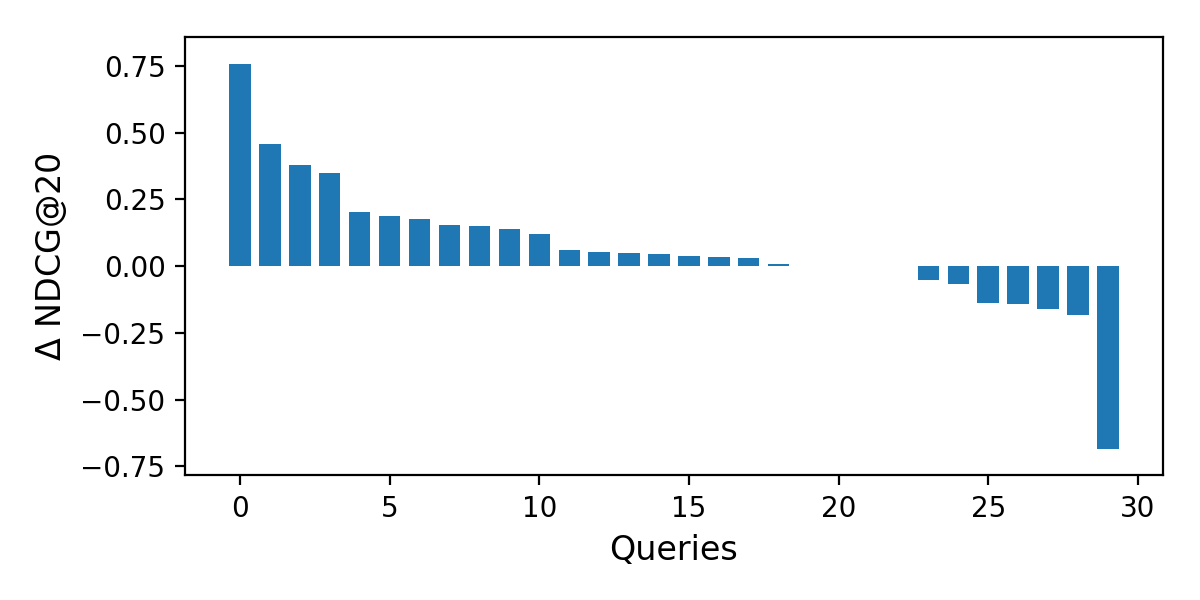}\label{fig:ndcg_diff_q2}}
	\end{tabular}
	\vspace*{-\baselineskip}
	\caption{Query-level differences on the two query subsets between the LTR baseline and STR. Positive values indicate improvements made by the latter.}
	 \label{fig:ndcg_diff_qs}
\end{figure*}

We plot the difference between the LTR baseline and STR for the two query subsets in Figure~\ref{fig:ndcg_diff_qs}.  Table~\ref{tbl:individual_queries} lists the queries that we discuss below.
The leftmost bar in Figure~\ref{fig:ndcg_diff_q1} corresponds to the query ``\emph{stocks}.'' For this broad query, there are two relevant and one highly relevant tables. LTR does not retrieve any highly relevant tables in the top 20, while STR manages to return one highly relevant table in the top 10.
The rightmost bar in Figure~\ref{fig:ndcg_diff_q1} corresponds to the query ``\emph{ibanez guitars}.'' For this query, there are two relevant and one highly relevant tables.  LTR produces an almost perfect ranking for this query, by returning the highly relevant table at the top rank, and the two relevant tables at ranks 2 and 4.  STR returns a non-relevant table at the top rank, thereby pushing the relevant results down in the ranking by a single position, resulting in a decrease of 0.29 in NDCG@20.

The leftmost bar in Figure~\ref{fig:ndcg_diff_q2} corresponds to the query ``\emph{board games number of players}.'' For this query, there are only two relevant tables according to the ground truth. STR managed to place them in the 1st and 3rd rank positions, while LTR returned only one of them at position 13th.
The rightmost bar in Figure~\ref{fig:ndcg_diff_q2} is the query ``\emph{cereals nutritional value}.'' Here, there is only one highly relevant result. LTR managed to place it in rank one, while it is ranked eighth by STR.
Another interesting query is ``\emph{irish counties area}'' (third bar from the left in Figure~\ref{fig:ndcg_diff_q2}), with three highly relevant and three relevant results according to the ground truth.  LTR returned two highly relevant and one relevant results at ranks 1, 2, and 4. STR, on the other hand, placed the three highly relevant results in the top 3 positions and also returned the three relevant tables at positions 4, 6, and 7.

%
%
\begin{table}[t]
\centering
\caption{Example queries from our query set. Rel denotes table relevance level. LTR and STR refer to the positions on which the table is returned by the respective method. }
\vspace*{-0.75\baselineskip}
\footnotesize
\begin{tabular}{lc@{~}c@{~}c@{~}}
	\toprule
	\textbf{Query} & Rel & LTR & STR  \\
	\midrule
	QS-1-24: \emph{stocks} \\
	~~~ Stocks for the Long Run / Key Data Findings: annual real returns & 2 & - & 6  \\
	~~~ TOPIX / TOPIX New Index Series & 1 & 9 & - \\
	~~~ Hang Seng Index / Selection criteria for the HSI constituent stocks & 1 & - & -  \\
	\midrule
	QS-1-21: \emph{ibanez guitars} \\ 
	 ~~~ Ibanez / Serial numbers & 2 & 1 & 2\\
	 ~~~ Corey Taylor / Equipment & 1 & 2 & 3 \\
	 ~~~ Fingerboard / Examples & 1 & 4 & 5 \\
	\midrule
	QS-2-27: \emph{board games number of players} \\
	~~~ List of Japanese board games & 1 & 13 & 1 \\ 
	 ~~~ List of licensed Risk game boards / Risk Legacy & 1 & - & 3 \\
	 
	 \midrule
	QS-2-21: \emph{cereals nutritional value} \\
	 ~~~ Sesame / Sesame seed kernels, toasted & 2 & 1 & 8 \\
	\midrule
	QS-2-20: \emph{irish counties area} \\
	~~~ Counties of Ireland / List of counties & 2 & 2 & 1 \\
	 ~~~ List of Irish counties by area / See also & 2 & 1 & 2 \\
	 ~~~ List of flags of Ireland / Counties of Ireland Flags & 2 & - & 3 \\
	 
	 ~~~ Provinces of Ireland / Demographics and politics & 1 & 4 & 4 \\
	~~~ Toponymical list of counties of the United Kingdom / Northern \dots & 1 & - & 7 \\
	~~~ Múscraige / Notes & 1 & - & 6 \\
	\bottomrule
\end{tabular}
\label{tbl:individual_queries}
\end{table}

\newpage
\section{Related Work}
There is an increasing amount of work on tables, addressing a wide range of tasks, including table search, table mining, table extension, and table completion.  Table search is a fundamental problem on its own, as well as used often as a core component in other tasks. 

\paragraph{Table Search}
Users are likely to search for tables when they need structured or relational data. 
\citet{Cafarella:2008:WEP} pioneered the table search task by introducing the WebTables system.  
The basic idea is to fetch the top-ranked results returned by a web search engine in response to the query, and then extract the top-$k$ tables from those pages.  Further refinements to the same idea are introduced in~\cite{Cafarella:2009:DIR}.
\citet{Venetis:2011:RST} leverage a database of class labels and relationships extracted from the Web, which are attached to table columns, for recovering table semantics. This information is then used to enhance table search.  \citet{Pimplikar:2012:ATQ} search for tables using column keywords, and match these keywords against the header, body, and context of tables. Google Web Tables\footnote{https://research.google.com/tables} provides an example of a table search system interface; 
the developers' experiences are summarized in~\cite{Balakrishnan:2015:AWP}. To enrich the diversity of search results, \citet{Nguyen:2015:RSS} design a goodness measure for table search and selection.  Apart from keyword-based search, tables may also be retrieved using a given ``local'' table as the query~\cite{Ahmadov:2015:THI,DasSarma:2012:FRT,Limaye:2010:ASW}.  We are not aware of any work that performs semantic matching of tables against queries.

\paragraph{Table Extension/Completion} 
\emph{Table extension} refers to the task of extending a table with additional elements, which are typically new columns~\cite{DasSarma:2012:FRT,Cafarella:2009:DIR,Lehmberg:2015:MSJ,Yakout:2012:IEA,Bhagavatula:2013:MEM}. These methods commonly use table search as the first step~\cite{Lehmberg:2015:MSJ,Bhagavatula:2013:MEM,Yakout:2012:IEA}. 
Searching related tables is also used for row extension. In~\cite{DasSarma:2012:FRT}, two tasks of entity complement and schema complement are addressed, to extend entity rows and columns respectively.  \citet{Zhang:2017:ESA} populate row and column headings of tables that have an entity focus. 
\emph{Table completion} is the task of filling in empty cells within a table. \citet{Ahmadov:2015:THI} introduce a method to extract table values from related tables and/or to predict them using machine learning methods.

\paragraph{Table Mining} 
The abundance of information in tables has raised great interest in table mining research~\cite{Cafarella:2011:SDW,Cafarella:2008:WEP,JM:2009:HDW,Sarawagi:2014:OQQ,Venetis:2011:RST,Zhang:2013:ISM}. 
\citet{Munoz:2014:ULD} recover table semantics by extracting RDF triples from Wikipedia tables. Similarly, \citet{Cafarella:2008:WEP} mine table relations from a huge table corpus extracted from a Google crawl.  Tables could also be searched to answer questions or mined to extend knowledge bases. \citet{Yin:2016:NEL} take tables as a knowledge base to execute queries using deep neural networks. \citet{Sekhavat:2014:KBA} augment an existing knowledge base (YAGO) with a probabilistic method by making use of table information. Similar work is carried out in~\cite{Dong:2014:KVW}, with tabular information used for knowledge base augmentation. Another line of work concerns table annotation and classification. \citet{Zwicklbauer:2013:TDW} introduce a method to annotate table headers by mining column content. \citet{Crestan:2011:WTC} introduce a supervised framework for classifying HTML tables into a taxonomy by examining the contents of a large number of tables. Apart from all the mentioned methods above, table mining also includes tasks like \emph{table interpretation}~\cite{Cafarella:2008:WEP, Munoz:2014:ULD, Venetis:2011:RST} and \emph{table recognition}~\cite{Crestan:2011:WTC,Zwicklbauer:2013:TDW}.  In the problem space of table mining, table search is an essential component.

\section{Conclusion}

In this paper, we have introduced and addressed the problem of ad hoc table retrieval: answering a keyword query with a ranked list of tables. 
We have developed a novel semantic matching framework, where queries and tables can be represented using semantic concepts (bag-of-entities and bag-of-categories) as well as continuous dense vectors (word and graph embeddings) in a uniform way. We have introduced multiple similarity measures for matching those semantic representations.
For evaluation, we have used a purpose-built test collection based on Wikipedia tables.  Finally, we have demonstrated substantial and significant improvements over a strong baseline. 
In future work, we wish to relax our requirements regarding the focus on Wikipedia tables, and make our methods applicable to other types of tables, like scientific tables~\citep{Gao:2017:STS} or Web tables.

\bibliographystyle{ACM-Reference-Format}
\balance
\bibliography{www2018-adhoc-table}


\begin{thebibliography}{00}


\ifx \showCODEN    \undefined \def \showCODEN     #1{\unskip}     \fi
\ifx \showDOI      \undefined \def \showDOI       #1{{\tt DOI:}\penalty0{#1}\ }
  \fi
\ifx \showISBNx    \undefined \def \showISBNx     #1{\unskip}     \fi
\ifx \showISBNxiii \undefined \def \showISBNxiii  #1{\unskip}     \fi
\ifx \showISSN     \undefined \def \showISSN      #1{\unskip}     \fi
\ifx \showLCCN     \undefined \def \showLCCN      #1{\unskip}     \fi
\ifx \shownote     \undefined \def \shownote      #1{#1}          \fi
\ifx \showarticletitle \undefined \def \showarticletitle #1{#1}   \fi
\ifx \showURL      \undefined \def \showURL       #1{#1}          \fi
\providecommand\bibfield[2]{#2}
\providecommand\bibinfo[2]{#2}
\providecommand\natexlab[1]{#1}
\providecommand\showeprint[2][]{arXiv:#2}

\bibitem[\protect\citeauthoryear{Ahmadov, Thiele, Eberius, Lehner, and
  Wrembel}{Ahmadov et~al\mbox{.}}{2015}]%
        {Ahmadov:2015:THI}
\bibfield{author}{\bibinfo{person}{Ahmad Ahmadov}, \bibinfo{person}{Maik
  Thiele}, \bibinfo{person}{Julian Eberius}, \bibinfo{person}{Wolfgang Lehner},
  {and} \bibinfo{person}{Robert Wrembel}.} \bibinfo{year}{2015}\natexlab{}.
\newblock \showarticletitle{Towards a Hybrid Imputation Approach Using Web
  Tables.}. In \bibinfo{booktitle}{{\em Proc. of BDC '15}}.
  \bibinfo{pages}{21--30}.
\newblock


\bibitem[\protect\citeauthoryear{Balakrishnan, Halevy, Harb, Lee, Madhavan,
  Rostamizadeh, Shen, Wilder, Wu, and Yu}{Balakrishnan et~al\mbox{.}}{2015}]%
        {Balakrishnan:2015:AWP}
\bibfield{author}{\bibinfo{person}{Sreeram Balakrishnan},
  \bibinfo{person}{Alon~Y. Halevy}, \bibinfo{person}{Boulos Harb},
  \bibinfo{person}{Hongrae Lee}, \bibinfo{person}{Jayant Madhavan},
  \bibinfo{person}{Afshin Rostamizadeh}, \bibinfo{person}{Warren Shen},
  \bibinfo{person}{Kenneth Wilder}, \bibinfo{person}{Fei Wu}, {and}
  \bibinfo{person}{Cong Yu}.} \bibinfo{year}{2015}\natexlab{}.
\newblock \showarticletitle{Applying WebTables in Practice}. In
  \bibinfo{booktitle}{{\em Proc. of CIDR '15}}.
\newblock


\bibitem[\protect\citeauthoryear{Balog, Bron, and De~Rijke}{Balog
  et~al\mbox{.}}{2011}]%
        {Balog:2011:QME}
\bibfield{author}{\bibinfo{person}{Krisztian Balog}, \bibinfo{person}{Marc
  Bron}, {and} \bibinfo{person}{Maarten De~Rijke}.}
  \bibinfo{year}{2011}\natexlab{}.
\newblock \showarticletitle{Query modeling for entity search based on terms,
  categories, and examples}.
\newblock \bibinfo{journal}{{\em ACM Trans. Inf. Syst.\/}}
  \bibinfo{volume}{29}, \bibinfo{number}{4}, Article \bibinfo{articleno}{22}
  (\bibinfo{date}{Dec.} \bibinfo{year}{2011}),
  \bibinfo{numpages}{22:1--22:31}~pages.
\newblock


\bibitem[\protect\citeauthoryear{Bhagavatula, Noraset, and Downey}{Bhagavatula
  et~al\mbox{.}}{2013}]%
        {Bhagavatula:2013:MEM}
\bibfield{author}{\bibinfo{person}{Chandra~Sekhar Bhagavatula},
  \bibinfo{person}{Thanapon Noraset}, {and} \bibinfo{person}{Doug Downey}.}
  \bibinfo{year}{2013}\natexlab{}.
\newblock \showarticletitle{Methods for Exploring and Mining Tables on
  Wikipedia}. In \bibinfo{booktitle}{{\em Proc. of IDEA '13}}.
  \bibinfo{pages}{18--26}.
\newblock


\bibitem[\protect\citeauthoryear{Bhagavatula, Noraset, and Downey}{Bhagavatula
  et~al\mbox{.}}{2015}]%
        {Bhagavatula:2015:TEL}
\bibfield{author}{\bibinfo{person}{Chandra~Sekhar Bhagavatula},
  \bibinfo{person}{Thanapon Noraset}, {and} \bibinfo{person}{Doug Downey}.}
  \bibinfo{year}{2015}\natexlab{}.
\newblock \showarticletitle{TabEL: Entity Linking in Web Tables}. In
  \bibinfo{booktitle}{{\em Proc. of ISWC '15}}. \bibinfo{pages}{425--441}.
\newblock


\bibitem[\protect\citeauthoryear{Cafarella, Halevy, and Khoussainova}{Cafarella
  et~al\mbox{.}}{2009}]%
        {Cafarella:2009:DIR}
\bibfield{author}{\bibinfo{person}{Michael~J. Cafarella}, \bibinfo{person}{Alon
  Halevy}, {and} \bibinfo{person}{Nodira Khoussainova}.}
  \bibinfo{year}{2009}\natexlab{}.
\newblock \showarticletitle{Data Integration for the Relational Web}.
\newblock \bibinfo{journal}{{\em Proc. of VLDB Endow.\/}}  \bibinfo{volume}{2}
  (\bibinfo{year}{2009}), \bibinfo{pages}{1090--1101}.
\newblock


\bibitem[\protect\citeauthoryear{Cafarella, Halevy, and Madhavan}{Cafarella
  et~al\mbox{.}}{2011}]%
        {Cafarella:2011:SDW}
\bibfield{author}{\bibinfo{person}{Michael~J. Cafarella}, \bibinfo{person}{Alon
  Halevy}, {and} \bibinfo{person}{Jayant Madhavan}.}
  \bibinfo{year}{2011}\natexlab{}.
\newblock \showarticletitle{Structured Data on the Web}.
\newblock \bibinfo{journal}{{\em Commun. ACM\/}}  \bibinfo{volume}{54}
  (\bibinfo{year}{2011}), \bibinfo{pages}{72--79}.
\newblock


\bibitem[\protect\citeauthoryear{Cafarella, Halevy, Wang, Wu, and
  Zhang}{Cafarella et~al\mbox{.}}{2008a}]%
        {Cafarella:2008:URW}
\bibfield{author}{\bibinfo{person}{Michael~J. Cafarella}, \bibinfo{person}{Alon
  Halevy}, \bibinfo{person}{Daisy~Zhe Wang}, \bibinfo{person}{Eugene Wu}, {and}
  \bibinfo{person}{Yang Zhang}.} \bibinfo{year}{2008}\natexlab{a}.
\newblock \showarticletitle{Uncovering the Relational Web}. In
  \bibinfo{booktitle}{{\em Proc. of WebDB '08}}.
\newblock


\bibitem[\protect\citeauthoryear{Cafarella, Halevy, Wang, Wu, and
  Zhang}{Cafarella et~al\mbox{.}}{2008b}]%
        {Cafarella:2008:WEP}
\bibfield{author}{\bibinfo{person}{Michael~J. Cafarella}, \bibinfo{person}{Alon
  Halevy}, \bibinfo{person}{Daisy~Zhe Wang}, \bibinfo{person}{Eugene Wu}, {and}
  \bibinfo{person}{Yang Zhang}.} \bibinfo{year}{2008}\natexlab{b}.
\newblock \showarticletitle{WebTables: Exploring the Power of Tables on the
  Web}.
\newblock \bibinfo{journal}{{\em Proc. of VLDB Endow.\/}}  \bibinfo{volume}{1}
  (\bibinfo{year}{2008}), \bibinfo{pages}{538--549}.
\newblock


\bibitem[\protect\citeauthoryear{Chen, Xiong, and Callan}{Chen
  et~al\mbox{.}}{2016}]%
        {Chen:2016:ESL}
\bibfield{author}{\bibinfo{person}{Jing Chen}, \bibinfo{person}{Chenyan Xiong},
  {and} \bibinfo{person}{Jamie Callan}.} \bibinfo{year}{2016}\natexlab{}.
\newblock \showarticletitle{An Empirical Study of Learning to Rank for Entity
  Search}. In \bibinfo{booktitle}{{\em Proc. of SIGIR '16}}.
  \bibinfo{pages}{737--740}.
\newblock


\bibitem[\protect\citeauthoryear{Crestan and Pantel}{Crestan and
  Pantel}{2011}]%
        {Crestan:2011:WTC}
\bibfield{author}{\bibinfo{person}{Eric Crestan} {and} \bibinfo{person}{Patrick
  Pantel}.} \bibinfo{year}{2011}\natexlab{}.
\newblock \showarticletitle{Web-scale Table Census and Classification}. In
  \bibinfo{booktitle}{{\em Proc. of WSDM '11}}. \bibinfo{pages}{545--554}.
\newblock


\bibitem[\protect\citeauthoryear{Das~Sarma, Fang, Gupta, Halevy, Lee, Wu, Xin,
  and Yu}{Das~Sarma et~al\mbox{.}}{2012}]%
        {DasSarma:2012:FRT}
\bibfield{author}{\bibinfo{person}{Anish Das~Sarma}, \bibinfo{person}{Lujun
  Fang}, \bibinfo{person}{Nitin Gupta}, \bibinfo{person}{Alon Halevy},
  \bibinfo{person}{Hongrae Lee}, \bibinfo{person}{Fei Wu},
  \bibinfo{person}{Reynold Xin}, {and} \bibinfo{person}{Cong Yu}.}
  \bibinfo{year}{2012}\natexlab{}.
\newblock \showarticletitle{Finding Related Tables}. In
  \bibinfo{booktitle}{{\em Proc. of SIGMOD '12}}. \bibinfo{pages}{817--828}.
\newblock


\bibitem[\protect\citeauthoryear{Dong, Gabrilovich, Heitz, Horn, Lao, Murphy,
  Strohmann, Sun, and Zhang}{Dong et~al\mbox{.}}{2014}]%
        {Dong:2014:KVW}
\bibfield{author}{\bibinfo{person}{Xin Dong}, \bibinfo{person}{Evgeniy
  Gabrilovich}, \bibinfo{person}{Geremy Heitz}, \bibinfo{person}{Wilko Horn},
  \bibinfo{person}{Ni Lao}, \bibinfo{person}{Kevin Murphy},
  \bibinfo{person}{Thomas Strohmann}, \bibinfo{person}{Shaohua Sun}, {and}
  \bibinfo{person}{Wei Zhang}.} \bibinfo{year}{2014}\natexlab{}.
\newblock \showarticletitle{Knowledge Vault: A Web-scale Approach to
  Probabilistic Knowledge Fusion}. In \bibinfo{booktitle}{{\em Proc. of KDD
  '14}}. \bibinfo{pages}{601--610}.
\newblock


\bibitem[\protect\citeauthoryear{Fleiss et~al\mbox{.}}{Fleiss
  et~al\mbox{.}}{1971}]%
        {Fleiss:1971:MNS}
\bibfield{author}{\bibinfo{person}{J.L. Fleiss} {and}
  \bibinfo{person}{others}.} \bibinfo{year}{1971}\natexlab{}.
\newblock \showarticletitle{{Measuring nominal scale agreement among many
  raters}}.
\newblock \bibinfo{journal}{{\em Psychological Bulletin\/}}
  \bibinfo{volume}{76} (\bibinfo{year}{1971}), \bibinfo{pages}{378--382}.
\newblock


\bibitem[\protect\citeauthoryear{Ganguly, Roy, Mitra, and Jones}{Ganguly
  et~al\mbox{.}}{2015}]%
        {Ganguly:2015:WEB}
\bibfield{author}{\bibinfo{person}{Debasis Ganguly}, \bibinfo{person}{Dwaipayan
  Roy}, \bibinfo{person}{Mandar Mitra}, {and} \bibinfo{person}{Gareth~J.F.
  Jones}.} \bibinfo{year}{2015}\natexlab{}.
\newblock \showarticletitle{Word Embedding Based Generalized Language Model for
  Information Retrieval}. In \bibinfo{booktitle}{{\em Proc. of SIGIR '15}}.
  \bibinfo{pages}{795--798}.
\newblock


\bibitem[\protect\citeauthoryear{Gao and Callan}{Gao and Callan}{2017}]%
        {Gao:2017:STS}
\bibfield{author}{\bibinfo{person}{Kyle~Yingkai Gao} {and}
  \bibinfo{person}{Jamie Callan}.} \bibinfo{year}{2017}\natexlab{}.
\newblock \showarticletitle{Scientific Table Search Using Keyword Queries}.
\newblock \bibinfo{journal}{{\em CoRR\/}}  \bibinfo{volume}{abs/1707.03423}
  (\bibinfo{year}{2017}).
\newblock


\bibitem[\protect\citeauthoryear{Grbovic, Djuric, Radosavljevic, Silvestri, and
  Bhamidipati}{Grbovic et~al\mbox{.}}{2015}]%
        {Grbovic:2015:CCE}
\bibfield{author}{\bibinfo{person}{Mihajlo Grbovic}, \bibinfo{person}{Nemanja
  Djuric}, \bibinfo{person}{Vladan Radosavljevic}, \bibinfo{person}{Fabrizio
  Silvestri}, {and} \bibinfo{person}{Narayan Bhamidipati}.}
  \bibinfo{year}{2015}\natexlab{}.
\newblock \showarticletitle{Context- and Content-aware Embeddings for Query
  Rewriting in Sponsored Search}. In \bibinfo{booktitle}{{\em Proc. of SIGIR
  '15}}. \bibinfo{pages}{383--392}.
\newblock


\bibitem[\protect\citeauthoryear{Hasibi, Balog, Garigliotti, and Zhang}{Hasibi
  et~al\mbox{.}}{2017a}]%
        {Hasibi:2017:NTE}
\bibfield{author}{\bibinfo{person}{Faegheh Hasibi}, \bibinfo{person}{Krisztian
  Balog}, \bibinfo{person}{Dar\'{\i}o Garigliotti}, {and} \bibinfo{person}{Shuo
  Zhang}.} \bibinfo{year}{2017}\natexlab{a}.
\newblock \showarticletitle{Nordlys: A Toolkit for Entity-Oriented and Semantic
  Search}. In \bibinfo{booktitle}{{\em Proceedings of SIGIR '17}}.
  \bibinfo{pages}{1289--1292}.
\newblock


\bibitem[\protect\citeauthoryear{Hasibi, Nikolaev, Xiong, Balog, Bratsberg,
  Kotov, and Callan}{Hasibi et~al\mbox{.}}{2017b}]%
        {Hasibi:2017:DVT}
\bibfield{author}{\bibinfo{person}{Faegheh Hasibi}, \bibinfo{person}{Fedor
  Nikolaev}, \bibinfo{person}{Chenyan Xiong}, \bibinfo{person}{Krisztian
  Balog}, \bibinfo{person}{Svein~Erik Bratsberg}, \bibinfo{person}{Alexander
  Kotov}, {and} \bibinfo{person}{Jamie Callan}.}
  \bibinfo{year}{2017}\natexlab{b}.
\newblock \showarticletitle{DBpedia-Entity V2: A Test Collection for Entity
  Search}. In \bibinfo{booktitle}{{\em Proc. of SIGIR '17}}.
  \bibinfo{pages}{1265--1268}.
\newblock


\bibitem[\protect\citeauthoryear{Kenter and de~Rijke}{Kenter and
  de~Rijke}{2015}]%
        {Kenter:2015:STS}
\bibfield{author}{\bibinfo{person}{Tom Kenter} {and} \bibinfo{person}{Maarten
  de Rijke}.} \bibinfo{year}{2015}\natexlab{}.
\newblock \showarticletitle{Short Text Similarity with Word Embeddings}. In
  \bibinfo{booktitle}{{\em Proc. of CIKM '15}}. \bibinfo{pages}{1411--1420}.
\newblock


\bibitem[\protect\citeauthoryear{Lehmberg, Ritze, Ristoski, Meusel, Paulheim,
  and Bizer}{Lehmberg et~al\mbox{.}}{2015}]%
        {Lehmberg:2015:MSJ}
\bibfield{author}{\bibinfo{person}{Oliver Lehmberg}, \bibinfo{person}{Dominique
  Ritze}, \bibinfo{person}{Petar Ristoski}, \bibinfo{person}{Robert Meusel},
  \bibinfo{person}{Heiko Paulheim}, {and} \bibinfo{person}{Christian Bizer}.}
  \bibinfo{year}{2015}\natexlab{}.
\newblock \showarticletitle{The Mannheim Search Join Engine}.
\newblock \bibinfo{journal}{{\em Web Semant.\/}}  \bibinfo{volume}{35}
  (\bibinfo{year}{2015}), \bibinfo{pages}{159--166}.
\newblock


\bibitem[\protect\citeauthoryear{Limaye, Sarawagi, and Chakrabarti}{Limaye
  et~al\mbox{.}}{2010}]%
        {Limaye:2010:ASW}
\bibfield{author}{\bibinfo{person}{Girija Limaye}, \bibinfo{person}{Sunita
  Sarawagi}, {and} \bibinfo{person}{Soumen Chakrabarti}.}
  \bibinfo{year}{2010}\natexlab{}.
\newblock \showarticletitle{Annotating and Searching Web Tables Using Entities,
  Types and Relationships}.
\newblock \bibinfo{journal}{{\em Proc. of VLDB Endow.\/}}  \bibinfo{volume}{3}
  (\bibinfo{year}{2010}), \bibinfo{pages}{1338--1347}.
\newblock


\bibitem[\protect\citeauthoryear{Liu}{Liu}{2011}]%
        {Liu:2011:LRI}
\bibfield{author}{\bibinfo{person}{Tie-Yan Liu}.}
  \bibinfo{year}{2011}\natexlab{}.
\newblock \bibinfo{booktitle}{{\em Learning to Rank for Information
  Retrieval}}.
\newblock \bibinfo{publisher}{Springer Berlin Heidelberg}.
\newblock


\bibitem[\protect\citeauthoryear{Macdonald, Santos, and Ounis}{Macdonald
  et~al\mbox{.}}{2012}]%
        {Macdonald:2012:UQF}
\bibfield{author}{\bibinfo{person}{Craig Macdonald}, \bibinfo{person}{Rodrygo
  L~T Santos}, {and} \bibinfo{person}{Iadh Ounis}.}
  \bibinfo{year}{2012}\natexlab{}.
\newblock \showarticletitle{On the Usefulness of Query Features for Learning to
  Rank}. In \bibinfo{booktitle}{{\em Proc. of CIKM '12}}.
  \bibinfo{pages}{2559--2562}.
\newblock


\bibitem[\protect\citeauthoryear{Madhavan, Afanasiev, Antova, and
  Halevy}{Madhavan et~al\mbox{.}}{2009}]%
        {JM:2009:HDW}
\bibfield{author}{\bibinfo{person}{Jayant Madhavan}, \bibinfo{person}{Loredana
  Afanasiev}, \bibinfo{person}{Lyublena Antova}, {and} \bibinfo{person}{Alon~Y.
  Halevy}.} \bibinfo{year}{2009}\natexlab{}.
\newblock \showarticletitle{Harnessing the Deep Web: Present and Future}.
\newblock \bibinfo{journal}{{\em CoRR\/}}  \bibinfo{volume}{abs/0909.1785}
  (\bibinfo{year}{2009}).
\newblock


\bibitem[\protect\citeauthoryear{Manotumruksa, MacDonald, and
  Ounis}{Manotumruksa et~al\mbox{.}}{2016}]%
        {Jarana:2016:MUP}
\bibfield{author}{\bibinfo{person}{Jarana Manotumruksa}, \bibinfo{person}{Craig
  MacDonald}, {and} \bibinfo{person}{Iadh Ounis}.}
  \bibinfo{year}{2016}\natexlab{}.
\newblock \showarticletitle{Modelling User Preferences using Word Embeddings
  for Context-Aware Venue Recommendation}.
\newblock \bibinfo{journal}{{\em CoRR\/}}  \bibinfo{volume}{abs/1606.07828}
  (\bibinfo{year}{2016}).
\newblock


\bibitem[\protect\citeauthoryear{Mikolov, Sutskever, Chen, Corrado, and
  Dean}{Mikolov et~al\mbox{.}}{2013}]%
        {Mikolov:2013:DRW}
\bibfield{author}{\bibinfo{person}{Tomas Mikolov}, \bibinfo{person}{Ilya
  Sutskever}, \bibinfo{person}{Kai Chen}, \bibinfo{person}{Greg Corrado}, {and}
  \bibinfo{person}{Jeffrey Dean}.} \bibinfo{year}{2013}\natexlab{}.
\newblock \showarticletitle{Distributed Representations of Words and Phrases
  and Their Compositionality}. In \bibinfo{booktitle}{{\em Proc. of NIPS '13}}.
  \bibinfo{pages}{3111--3119}.
\newblock


\bibitem[\protect\citeauthoryear{Mitra, Nalisnick, Craswell, and Caruana}{Mitra
  et~al\mbox{.}}{2016}]%
        {Bhaskar:2016:DES}
\bibfield{author}{\bibinfo{person}{Bhaskar Mitra}, \bibinfo{person}{Eric~T.
  Nalisnick}, \bibinfo{person}{Nick Craswell}, {and} \bibinfo{person}{Rich
  Caruana}.} \bibinfo{year}{2016}\natexlab{}.
\newblock \showarticletitle{A Dual Embedding Space Model for Document Ranking}.
\newblock \bibinfo{journal}{{\em CoRR\/}}  \bibinfo{volume}{abs/1602.01137}
  (\bibinfo{year}{2016}).
\newblock


\bibitem[\protect\citeauthoryear{Munoz, Hogan, and Mileo}{Munoz
  et~al\mbox{.}}{2014}]%
        {Munoz:2014:ULD}
\bibfield{author}{\bibinfo{person}{Emir Munoz}, \bibinfo{person}{Aidan Hogan},
  {and} \bibinfo{person}{Alessandra Mileo}.} \bibinfo{year}{2014}\natexlab{}.
\newblock \showarticletitle{Using Linked Data to Mine RDF from Wikipedia's
  Tables}. In \bibinfo{booktitle}{{\em Proc. of WSDM '14}}.
  \bibinfo{pages}{533--542}.
\newblock


\bibitem[\protect\citeauthoryear{Nguyen, Nguyen, Matthias, and Karl}{Nguyen
  et~al\mbox{.}}{2015}]%
        {Nguyen:2015:RSS}
\bibfield{author}{\bibinfo{person}{Thanh~Tam Nguyen}, \bibinfo{person}{Quoc
  Viet~Hung Nguyen}, \bibinfo{person}{Weidlich Matthias}, {and}
  \bibinfo{person}{Aberer Karl}.} \bibinfo{year}{2015}\natexlab{}.
\newblock \showarticletitle{Result Selection and Summarization for Web Table
  Search}. In \bibinfo{booktitle}{{\em ISDE '15}}. \bibinfo{pages}{231--242}.
\newblock


\bibitem[\protect\citeauthoryear{Ogilvie and Callan}{Ogilvie and
  Callan}{2003}]%
        {Ogilvie:2003:MLM}
\bibfield{author}{\bibinfo{person}{Paul Ogilvie} {and} \bibinfo{person}{Jamie
  Callan}.} \bibinfo{year}{2003}\natexlab{}.
\newblock \showarticletitle{Combining Document Representations for Known-item
  Search}. In \bibinfo{booktitle}{{\em Proc. of SIGIR '03}}.
  \bibinfo{pages}{143--150}.
\newblock


\bibitem[\protect\citeauthoryear{Pennington, Socher, and Manning}{Pennington
  et~al\mbox{.}}{2014}]%
        {Pennington:2014:GGV}
\bibfield{author}{\bibinfo{person}{Jeffrey Pennington},
  \bibinfo{person}{Richard Socher}, {and} \bibinfo{person}{Christopher~D
  Manning}.} \bibinfo{year}{2014}\natexlab{}.
\newblock \showarticletitle{{GloVe}: Global Vectors for Word Representation}.
  In \bibinfo{booktitle}{{\em Proc. of EMNLP '14}}.
  \bibinfo{pages}{1532--1543}.
\newblock


\bibitem[\protect\citeauthoryear{Perozzi, Al-Rfou, and Skiena}{Perozzi
  et~al\mbox{.}}{2014}]%
        {Perozzi:2014:DOL}
\bibfield{author}{\bibinfo{person}{Bryan Perozzi}, \bibinfo{person}{Rami
  Al-Rfou}, {and} \bibinfo{person}{Steven Skiena}.}
  \bibinfo{year}{2014}\natexlab{}.
\newblock \showarticletitle{DeepWalk: Online Learning of Social
  Representations}. In \bibinfo{booktitle}{{\em Proc. of KDD '14}}.
  \bibinfo{pages}{701--710}.
\newblock


\bibitem[\protect\citeauthoryear{Pimplikar and Sarawagi}{Pimplikar and
  Sarawagi}{2012}]%
        {Pimplikar:2012:ATQ}
\bibfield{author}{\bibinfo{person}{Rakesh Pimplikar} {and}
  \bibinfo{person}{Sunita Sarawagi}.} \bibinfo{year}{2012}\natexlab{}.
\newblock \showarticletitle{Answering Table Queries on the Web Using Column
  Keywords}.
\newblock \bibinfo{journal}{{\em Proc. of VLDB Endow.\/}}  \bibinfo{volume}{5}
  (\bibinfo{year}{2012}), \bibinfo{pages}{908--919}.
\newblock


\bibitem[\protect\citeauthoryear{Qin, Liu, Xu, and Li}{Qin
  et~al\mbox{.}}{2010}]%
        {Qin:2010:LBC}
\bibfield{author}{\bibinfo{person}{Tao Qin}, \bibinfo{person}{Tie-Yan Liu},
  \bibinfo{person}{Jun Xu}, {and} \bibinfo{person}{Hang Li}.}
  \bibinfo{year}{2010}\natexlab{}.
\newblock \showarticletitle{{LETOR}: A Benchmark Collection for Research on
  Learning to Rank for Information Retrieval}.
\newblock \bibinfo{journal}{{\em Inf. Retr.\/}} \bibinfo{volume}{13},
  \bibinfo{number}{4} (\bibinfo{date}{Aug} \bibinfo{year}{2010}),
  \bibinfo{pages}{346--374}.
\newblock


\bibitem[\protect\citeauthoryear{Raviv, Kurland, and Carmel}{Raviv
  et~al\mbox{.}}{2016}]%
        {Raviv:2016:DRU}
\bibfield{author}{\bibinfo{person}{Hadas Raviv}, \bibinfo{person}{Oren
  Kurland}, {and} \bibinfo{person}{David Carmel}.}
  \bibinfo{year}{2016}\natexlab{}.
\newblock \showarticletitle{Document Retrieval Using Entity-Based Language
  Models}. In \bibinfo{booktitle}{{\em Proc. of SIGIR '16}}.
  \bibinfo{pages}{65--74}.
\newblock


\bibitem[\protect\citeauthoryear{Ristoski and Paulheim}{Ristoski and
  Paulheim}{2016}]%
        {Ristoski:2016:RGE}
\bibfield{author}{\bibinfo{person}{Petar Ristoski} {and} \bibinfo{person}{Heiko
  Paulheim}.} \bibinfo{year}{2016}\natexlab{}.
\newblock \showarticletitle{{RDF2vec}: {RDF} Graph Embeddings for Data Mining}.
  In \bibinfo{booktitle}{{\em Proc. of ISWC '16}}. \bibinfo{pages}{498--514}.
\newblock


\bibitem[\protect\citeauthoryear{Sarawagi and Chakrabarti}{Sarawagi and
  Chakrabarti}{2014}]%
        {Sarawagi:2014:OQQ}
\bibfield{author}{\bibinfo{person}{Sunita Sarawagi} {and}
  \bibinfo{person}{Soumen Chakrabarti}.} \bibinfo{year}{2014}\natexlab{}.
\newblock \showarticletitle{Open-domain Quantity Queries on Web Tables:
  Annotation, Response, and Consensus Models}. In \bibinfo{booktitle}{{\em
  Proc. of KDD '14}}. \bibinfo{pages}{711--720}.
\newblock


\bibitem[\protect\citeauthoryear{Sekhavat, Paolo, Barbosa, and
  Merialdo}{Sekhavat et~al\mbox{.}}{2014}]%
        {Sekhavat:2014:KBA}
\bibfield{author}{\bibinfo{person}{Yoones~A. Sekhavat},
  \bibinfo{person}{Francesco~Di Paolo}, \bibinfo{person}{Denilson Barbosa},
  {and} \bibinfo{person}{Paolo Merialdo}.} \bibinfo{year}{2014}\natexlab{}.
\newblock \showarticletitle{Knowledge Base Augmentation using Tabular Data}. In
  \bibinfo{booktitle}{{\em Proc. of LDOW '14}}.
\newblock


\bibitem[\protect\citeauthoryear{Tang, Qu, Wang, Zhang, Yan, and Mei}{Tang
  et~al\mbox{.}}{2015}]%
        {Tang:2015:LLI}
\bibfield{author}{\bibinfo{person}{Jian Tang}, \bibinfo{person}{Meng Qu},
  \bibinfo{person}{Mingzhe Wang}, \bibinfo{person}{Ming Zhang},
  \bibinfo{person}{Jun Yan}, {and} \bibinfo{person}{Qiaozhu Mei}.}
  \bibinfo{year}{2015}\natexlab{}.
\newblock \showarticletitle{LINE: Large-scale Information Network Embedding}.
  In \bibinfo{booktitle}{{\em Proc. of WWW '15}}. \bibinfo{pages}{1067--1077}.
\newblock


\bibitem[\protect\citeauthoryear{Tyree, Weinberger, Agrawal, and Paykin}{Tyree
  et~al\mbox{.}}{2011}]%
        {Tyree:2011:PBR}
\bibfield{author}{\bibinfo{person}{Stephen Tyree}, \bibinfo{person}{Kilian~Q
  Weinberger}, \bibinfo{person}{Kunal Agrawal}, {and} \bibinfo{person}{Jennifer
  Paykin}.} \bibinfo{year}{2011}\natexlab{}.
\newblock \showarticletitle{Parallel Boosted Regression Trees for Web Search
  Ranking}. In \bibinfo{booktitle}{{\em Proc. of WWW '11}}.
  \bibinfo{pages}{387--396}.
\newblock


\bibitem[\protect\citeauthoryear{Venetis, Halevy, Madhavan, Pa\c{s}ca, Shen,
  Wu, Miao, and Wu}{Venetis et~al\mbox{.}}{2011}]%
        {Venetis:2011:RST}
\bibfield{author}{\bibinfo{person}{Petros Venetis}, \bibinfo{person}{Alon
  Halevy}, \bibinfo{person}{Jayant Madhavan}, \bibinfo{person}{Marius
  Pa\c{s}ca}, \bibinfo{person}{Warren Shen}, \bibinfo{person}{Fei Wu},
  \bibinfo{person}{Gengxin Miao}, {and} \bibinfo{person}{Chung Wu}.}
  \bibinfo{year}{2011}\natexlab{}.
\newblock \showarticletitle{Recovering Semantics of Tables on the Web}.
\newblock \bibinfo{journal}{{\em Proc. of VLDB Endow.\/}}  \bibinfo{volume}{4}
  (\bibinfo{year}{2011}), \bibinfo{pages}{528--538}.
\newblock


\bibitem[\protect\citeauthoryear{Vuli\'{c} and Moens}{Vuli\'{c} and
  Moens}{2015}]%
        {Vulic:2015:MCI}
\bibfield{author}{\bibinfo{person}{Ivan Vuli\'{c}} {and}
  \bibinfo{person}{Marie-Francine Moens}.} \bibinfo{year}{2015}\natexlab{}.
\newblock \showarticletitle{Monolingual and Cross-Lingual Information Retrieval
  Models Based on (Bilingual) Word Embeddings}. In \bibinfo{booktitle}{{\em
  Proc. of SIGIR '15}}. \bibinfo{pages}{363--372}.
\newblock


\bibitem[\protect\citeauthoryear{Xiong, Callan, and Liu}{Xiong
  et~al\mbox{.}}{2017}]%
        {Xiong:2017:WDR}
\bibfield{author}{\bibinfo{person}{Chenyan Xiong}, \bibinfo{person}{Jamie
  Callan}, {and} \bibinfo{person}{Tie-Yan Liu}.}
  \bibinfo{year}{2017}\natexlab{}.
\newblock \showarticletitle{Word-Entity Duet Representations for Document
  Ranking}. In \bibinfo{booktitle}{{\em Proc. of SIGIR '17}}.
  \bibinfo{pages}{763--772}.
\newblock


\bibitem[\protect\citeauthoryear{Yakout, Ganjam, Chakrabarti, and
  Chaudhuri}{Yakout et~al\mbox{.}}{2012}]%
        {Yakout:2012:IEA}
\bibfield{author}{\bibinfo{person}{Mohamed Yakout}, \bibinfo{person}{Kris
  Ganjam}, \bibinfo{person}{Kaushik Chakrabarti}, {and}
  \bibinfo{person}{Surajit Chaudhuri}.} \bibinfo{year}{2012}\natexlab{}.
\newblock \showarticletitle{InfoGather: Entity Augmentation and Attribute
  Discovery by Holistic Matching with Web Tables}. In \bibinfo{booktitle}{{\em
  Proc. of SIGMOD '12}}. \bibinfo{pages}{97--108}.
\newblock


\bibitem[\protect\citeauthoryear{Yin, Lu, Li, and Kao}{Yin
  et~al\mbox{.}}{2016}]%
        {Yin:2016:NEL}
\bibfield{author}{\bibinfo{person}{Pengcheng Yin}, \bibinfo{person}{Zhengdong
  Lu}, \bibinfo{person}{Hang Li}, {and} \bibinfo{person}{Ben Kao}.}
  \bibinfo{year}{2016}\natexlab{}.
\newblock \showarticletitle{Neural Enquirer: Learning to Query Tables in
  Natural Language}. In \bibinfo{booktitle}{{\em Proc. of IJCAI '16}}.
  \bibinfo{pages}{2308--2314}.
\newblock


\bibitem[\protect\citeauthoryear{Zhang and Chakrabarti}{Zhang and
  Chakrabarti}{2013}]%
        {Zhang:2013:ISM}
\bibfield{author}{\bibinfo{person}{Meihui Zhang} {and} \bibinfo{person}{Kaushik
  Chakrabarti}.} \bibinfo{year}{2013}\natexlab{}.
\newblock \showarticletitle{InfoGather+: Semantic Matching and Annotation of
  Numeric and Time-varying Attributes in Web Tables}. In
  \bibinfo{booktitle}{{\em Proc. of SIGMOD '13}}. \bibinfo{pages}{145--156}.
\newblock


\bibitem[\protect\citeauthoryear{Zhang and Balog}{Zhang and Balog}{2017a}]%
        {Shuo:2017:DPF}
\bibfield{author}{\bibinfo{person}{Shuo Zhang} {and} \bibinfo{person}{Krisztian
  Balog}.} \bibinfo{year}{2017}\natexlab{a}.
\newblock \showarticletitle{Design Patterns for Fusion-Based Object Retrieval}.
  In \bibinfo{booktitle}{{\em Proc. of ECIR '17}}. \bibinfo{pages}{684--690}.
\newblock


\bibitem[\protect\citeauthoryear{Zhang and Balog}{Zhang and Balog}{2017b}]%
        {Zhang:2017:ESA}
\bibfield{author}{\bibinfo{person}{Shuo Zhang} {and} \bibinfo{person}{Krisztian
  Balog}.} \bibinfo{year}{2017}\natexlab{b}.
\newblock \showarticletitle{EntiTables: Smart Assistance for Entity-Focused
  Tables}. In \bibinfo{booktitle}{{\em Proc. of SIGIR '17}}.
  \bibinfo{pages}{255--264}.
\newblock


\bibitem[\protect\citeauthoryear{Zhou, He, Zhao, and Hu}{Zhou
  et~al\mbox{.}}{2015}]%
        {zhou-EtAl:2015:ACL-IJCNLP1}
\bibfield{author}{\bibinfo{person}{Guangyou Zhou}, \bibinfo{person}{Tingting
  He}, \bibinfo{person}{Jun Zhao}, {and} \bibinfo{person}{Po Hu}.}
  \bibinfo{year}{2015}\natexlab{}.
\newblock \showarticletitle{Learning Continuous Word Embedding with Metadata
  for Question Retrieval in Community Question Answering}. In
  \bibinfo{booktitle}{{\em Proc. of ACL '15}}. \bibinfo{pages}{250--259}.
\newblock


\bibitem[\protect\citeauthoryear{Zwicklbauer, Einsiedler, Granitzer, and
  Seifert}{Zwicklbauer et~al\mbox{.}}{2013}]%
        {Zwicklbauer:2013:TDW}
\bibfield{author}{\bibinfo{person}{Stefan Zwicklbauer},
  \bibinfo{person}{Christoph Einsiedler}, \bibinfo{person}{Michael Granitzer},
  {and} \bibinfo{person}{Christin Seifert}.} \bibinfo{year}{2013}\natexlab{}.
\newblock \showarticletitle{Towards Disambiguating Web Tables}. In
  \bibinfo{booktitle}{{\em Proc. of ISWC-PD '13}}. \bibinfo{pages}{205--208}.
\newblock


\end{thebibliography}

\end{document}